\documentclass[12pt, a4paper]{article}
\renewcommand{\baselinestretch}{1.5}
\usepackage[margin=1in]{geometry}
\usepackage{graphicx} 
\usepackage{float}
\usepackage[font=footnotesize]{caption}
\usepackage{hyperref}
\usepackage{orcidlink}
\usepackage{booktabs}
\usepackage[most]{tcolorbox}
\usepackage[dvipsnames]{xcolor}
\usepackage{enumitem}
\usepackage{caption}
\usepackage{float}
\usepackage{tabularx}
\usepackage{booktabs}
\usepackage[backend=biber, style=numeric-comp,
            sorting=none]{biblatex}
\let\cite\supercite 

\title{\sffamily\bfseries  
Hierarchical Multi-agent Large Language Model Reasoning for Autonomous Functional Materials Discovery
}

\author{\vspace{-5ex}}
\date{\vspace{-5ex}}

\begin{document}

\maketitle

\vspace{-8ex}

\begin{center}
    \large \sffamily
    Samuel Rothfarb$^{\dagger,\ddagger}$ \orcidlink{0009-0004-6696-9495}, 
    Megan C. Davis$^\ddagger$ \orcidlink{0000-0002-4038-8615}, 
    Ivana Matanovic$^\ddagger$ \orcidlink{0000-0002-9191-8620}, \\
    \large \sffamily
    Baikun Li$^\dagger$* \orcidlink{0000-0002-5623-5912}, 
    Edward F. Holby$^\ddagger$* \orcidlink{0000-0001-8419-6298}, and Wilton J.M. Kort-Kamp$^{\ddagger}$* \orcidlink{0000-0002-0679-6690}\\
\end{center}

\begin{center}
    \normalsize 
    \textit{
    \noindent $^\dagger$School of Civil \& Environmental Engineering, University of Connecticut, Storrs, Connecticut 06269, United States.\\
    \noindent $^\ddagger$Theoretical Division, Los Alamos National Laboratory, Los Alamos, New Mexico 87545, United States.\\   }
\end{center}

\begin{center}
    \sffamily
    *Corresponding authors: \href{baikun.li@uconn.edu}{baikun.li@uconn.edu}, \href{holby@lanl.gov}{holby@lanl.gov}, \href{mailto:kortkamp@lanl.gov}{kortkamp@lanl.gov}\\
\end{center}

\vspace{2ex}

\begin{abstract} 
    Artificial intelligence is reshaping scientific exploration, but most methods automate procedural tasks without engaging in scientific reasoning, limiting autonomy in discovery. We introduce Materials Agents for Simulation and Theory in Electronic-structure Reasoning (MASTER), an active learning framework where large language models autonomously design, execute, and interpret atomistic simulations. In MASTER, a multimodal system translates natural language into density functional theory workflows, while higher-level reasoning agents guide discovery through a hierarchy of strategies, including a single agent baseline and three multi-agent approaches: peer review, triage-ranking, and triage-forms. Across two chemical applications, CO adsorption on Cu-surface transition metal (M) adatoms and on M–N–C catalysts, reasoning-driven exploration reduces required atomistic simulations by up to 90\% relative to trial-and-error selection. Reasoning trajectories reveal chemically grounded decisions that cannot be explained by stochastic sampling or semantic bias. Altogether, multi-agent collaboration accelerates materials discovery and marks a new paradigm for autonomous scientific exploration.
\end{abstract}
\vspace{30pt}
\begin{tabular}{@{}l@{\,}p{0.8\linewidth}}
\textbf{Keywords:} &
\textit{Multi-agent reasoning; Large language models; Active learning;
AI-driven simulation; Materials discovery; Density functional theory;
Surface chemistry.}
\end{tabular}
\vspace{30pt}
\section*{Introduction}


Recent advances in artificial intelligence (AI) have expanded its role in scientific research, enabling systems to analyze data, identify patterns, and even propose hypotheses. \cite{BRANDA2025101727} Yet, most of these models operate within fixed objectives and have limited ability to deliberate about scientific questions or adapt their strategies based on outcomes. Reasoning, the process of evaluating competing hypotheses and designing informative experiments, is central to genuine scientific exploration. In materials research, where the search space spans millions of possible atomic configurations, \cite{MAGUS, Chen_2024, LENG2025113775} such reasoning ability could enable targeted efforts toward the most informative regions of chemical space, transforming the rate and nature of discovery. 
Despite this potential, reasoning-driven autonomy remains challenging in materials science. Materials design is resource-intensive, requiring specialized expertise and substantial computational effort. \cite{high_throughput_computational} Even with advances in automation, the process from concept to verified material can extend over years. \cite{zhong_accelerated_2020} Current high-throughput atomistic simulation workflows have accelerated certain aspects of this process by enabling parallel screening of large chemical spaces,\cite{pyiron-paper, annevelink_automat_2022} but they offer limited adaptivity and cannot reason to select the next computation. Failed calculations still require human intervention to diagnose and repair errors, limiting true automation.

Machine learning has introduced new paradigms for accelerating materials discovery through data-driven predictive and generative modeling. \cite{merchant_scaling_2023,MatterGen,wood2025umafamilyuniversalmodels} However, while these models achieve high accuracy, they remain confined to their training distributions and cannot interpret results within an adaptive, iterative scientific framework. Because materials discovery is inherently a multi-stage reasoning process, from hypothesis generation to experimental validation, these models play a crucial but limited role, addressing computation rather than scientific decision-making.

Large language models (LLMs) bring reasoning capabilities to scientific research. They integrate knowledge, generate executable code, interface with tools, and reason in natural language. \cite{vaswani2023attentionneed, devlin2019bertpretrainingdeepbidirectional, brown2020languagemodelsfewshotlearners} These abilities position LLMs as coordinators of end-to-end discovery workflows that link hypothesis generation, simulation, and analysis. However, LLMs acting in isolation face challenges including limited numerical precision, error accumulation, and lack of persistent state. \cite{MORADI20251681} Stateless interactions prevent them from retaining context, coordinating across tools, or adapting strategies based on intermediate results. Robust scientific reasoning therefore requires agentic architectures, \cite{LLMSurveyPaper} i.e., systems coupling LLMs to memory, adaptive planning, and iterative feedback. 
Recent studies have shown that LLM agents can manage computational materials workflows with impressive autonomy. Systems such as VASPilot,~\cite{VASPilot2025} DynaMate,~\cite{DynaMate2025} and ChemGraph,~\cite{ChemGraph2025} perform procedural automation quantum-mechanical simulations, with VASPilot enabling end-to-end execution of VASP~\cite{VASP1, VASP2, VASP3} workflows spanning structure preparation, job execution, error recovery, and postprocessing. Beyond single-job automation, DREAMS~\cite{DREAMS2025} and MOFGen~\cite{MOFGen2024} extend autonomy to networked simulation environments. In DREAMS, specialized agents coordinate the setup and execution of Quantum Espresso~\cite{Giannozzi_2009, Giannozzi_2017} calculations with automated convergence checks and recovery from failure. Moreover, LLMatDesign \cite{LLMatDesign2024} introduced a self-reflective loop in which a LLM refined its proposals using a surrogate model pre-trained on density functional theory (DFT) data.

Here, we introduce Materials Agents for Simulation and Theory in Electronic-structure Reasoning (MASTER), an active learning framework that equips ensembles of LLMs with structured, collective reasoning for guided materials discovery. Designed around reasoning autonomy, MASTER coordinates interacting agents to deliberate, critique, and refine hypotheses through structured collaboration, deciding what atomistic simulations to perform and how to interpret their outcomes. Within this framework (Fig. \ref{fig:our_system_and_materials_problem}a), a multimodal subsystem links natural-language objectives to validated DFT workflows, ensuring accurate generation of atomic structures, input files, and convergence parameters. Meanwhile, higher-level reasoning agents decide which materials to investigate next based on intermediate results. This separation of concerns allows the system to frame decision-making as a reasoning process that integrates individual perspectives into shared conclusions while preserving the rigor of first-principles computation.
To evaluate how reasoning architecture influences discovery efficiency, we compare a single-agent baseline with three hierarchical multi-agent strategies—peer review, triage-ranking, and triage-forms. These architectures are tested using CO binding energetics across two chemical domains:  transition-metal (M) adatoms on a Cu(100)  surface and M–N–C single-atom catalysts. \cite{ChungScience2017, liang_electrochemical_2021, DAVIS2025145357} MASTER identifies targeted binding energies within only a few iterations, whereas trial-and-error selection requires an order of magnitude more trials. For the highest-performing systems, analysis of reasoning trajectories reveals coherent and scientifically grounded decision patterns, reflecting how structured collaboration shapes adaptive, mechanistically informed exploration.

\section*{Results}

\subsection*{\textit{MASTER Framework and Benchmark Materials Problem}}

\begin{figure}[!t]
    \centering
    \includegraphics[width=\textwidth]{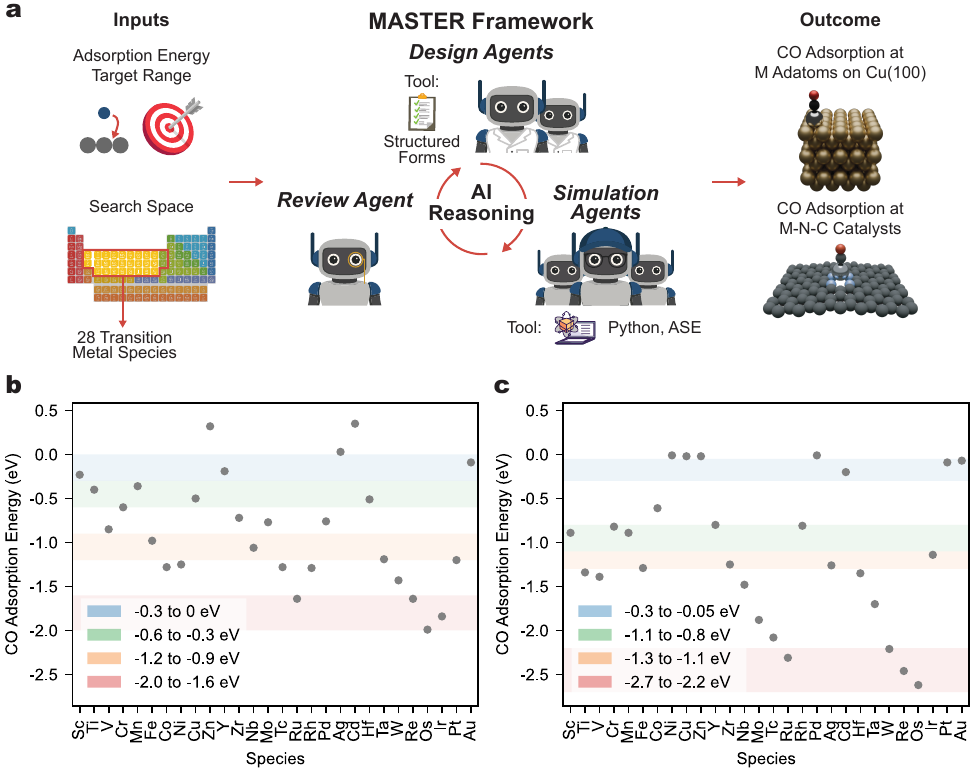}
    \caption{Unlocking autonomy in materials discovery via large language models-driven active learning. \textbf{a}, Schematic of the MASTER framework, where design agents propose materials, simulation agents generate atomic positions scripts, and a reviewer agent evaluates if outcomes meet required targets. \textbf{b}, Computed CO adsorption energies for transition-metal adatoms on Cu(100). \textbf{c}, Computed CO adsorption energies for M–N–C single-atom catalysts. In both \textbf{b} and \textbf{c} the shaded bands mark target energy windows from weak to strong binding. A comparison of the CO adsorption energies is presented in Figure \ref{fig:adatom_vs_mnc}.}
    \label{fig:our_system_and_materials_problem}
\end{figure}

The MASTER framework integrates LLM reasoning with autonomous electronic-structure simulation in a closed loop that emulates the operation of a scientific research team (Fig. \ref{fig:our_system_and_materials_problem}a). The system is organized into three tightly coupled yet functionally distinct layers. The design layer comprises a team of LLM agents that use natural language to formulate hypotheses and deliberate, individually or collectively, on which material to evaluate next based on the accumulated simulation history. Once a candidate structure is chosen, the information is passed to the simulation layer, which provides a multimodal interface that converts high-level simulation objectives into validated DFT workflows. Here, a team of DFT agents autonomously generates inputs, atomic geometries, and executes first-principles calculations. The computed quantities, such as adsorption energies, are then returned to the review layer where a reviewer agent determines whether the specified criteria have been met or further exploration is required. By explicitly separating reasoning from simulation, the design agents focus on interpreting trends, weighing evidence, and planning experiments, while the DFT and reviewer agents handle numerical execution and verification, respectively. Together, these layers operationalize an autonomous scientific method, transforming static computational screening into an adaptive and self-correcting exploration of chemical space through LLM reasoning and first-principles feedback. 

We apply MASTER to the problem of CO adsorption energetics, which has been widely studied in surface science and heterogeneous catalysis. \cite{roughening, probing_structural} CO binding energy serves as a key descriptor of catalytic activity and selectivity, \cite{VAZQUEZPARGA2023156581, Mesostructure} governing reaction pathways in CO oxidation, \cite{GAO2022832} CO$_2$ reduction, \cite{StuructureSensitivity} and the formation of C$_2$ products via CO dimerization. \cite{MechanisticInvestigation, SurfaceEngineering} It depends sensitively on the local electronic and geometric environment, producing a chemically intricate landscape that challenges autonomous LLM reasoning.  Two scenarios are considered here (Figs. \ref{fig:our_system_and_materials_problem}b and \ref{fig:our_system_and_materials_problem}c). The first examines CO adsorption on transition-metal adatoms supported on Cu(100), a model for undercoordinated catalytic sites where low coordination substantially modifies the local binding environment. \cite{christiansen_single-atom_2024} The second extends the study to CO adsorption on M-N-C single-atom catalysts, \cite{DAVIS2025145357} where the metal is coordinated by nitrogen ligands within a graphene host, introducing distinct ligand-field and covalency effects. Each domain spans twenty-eight transition metals from Sc to Au with DFT-computed adsorption energies. Four target CO binding-energy ranges were chosen in each case, spanning weak to strong binding regimes (Methods). In this context, MASTER must learn and generalize structure–property relationships directly from first-principles data and navigate complex search spaces without human guidance.

\subsection*{\textit{Natural Language to Density Functional Theory Simulations}}


The adatom adsorption problem provides an ideal proof-of-principle for MASTER because it represents a complex case that encapsulates the fundamental challenges in automating atomic-scale simulation, especially for the field of electrocatalysis.~\cite{DFT_surface_chemistry} Indeed, researchers may describe an idea succinctly, such as “place a CO molecule on an Ag adatom supported on Cu(100)”, but DFT codes require constructing the appropriate crystallographic surface, identifying high-symmetry adsorption sites, optimizing supercell dimensions, and defining vacuum spacing to avoid artificial slab couplings. MASTER must therefore translate scientific intent expressed in natural language, into fully executable simulations. Even when computational choices such as exchange–correlation functional or $k$-point meshes remain constant, each new atomic system must be built from scratch to capture the intended chemistry accurately. Thus, the Cu(100) adatom system offers a rigorous benchmark in which success or failure is unambiguous and the translation challenge is fully exposed.

\begin{figure}[t]
    \centering
    \includegraphics[width=\textwidth]{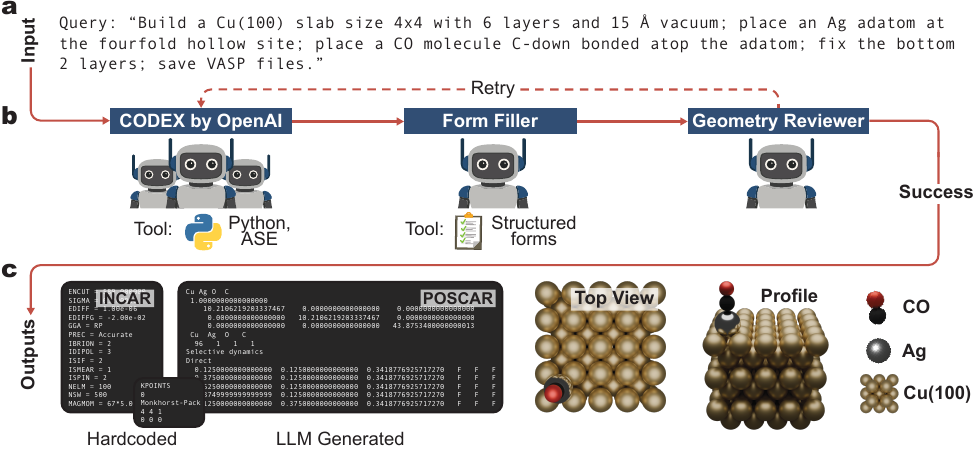}
    \caption{MASTER'S simulation agents convert natural-language into first-principles calculations. \textbf{a}, Example query specifying the construction of a CO–Ag adatom system on Cu(100). \textbf{b}, The DFT subsystem contains three collaborating agents that generate and verify atomic positions. CODEX \cite{chen2021evaluatinglargelanguagemodels} produces an initial atomic structure by writing code that constructs the geometry using ASE. \cite{ase-paper, ase-paper-2} The Form Filler then evaluates the VASP~\cite{VASP1, VASP2, VASP3} structure file (POSCAR) together with visualizations of the top, profile, and side views. It completes an expert-prepared structured form (SI Note \ref{note:geo_review}) that focuses the agent’s reasoning context on the most common fault points in surface geometries. The Geometry Reviewer reads this form to determine whether the geometry is correct. If it is incorrect, then it issues a retry request with targeted feedback to CODEX. This loop continues until the geometry satisfies all criteria. \textbf{c}, The final output is the validated POSCAR file describing the intended adsorbate-surface configuration.}
    
    \label{fig:lang_to_DFT}
\end{figure}

To operationalize this translation, the simulation layer of MASTER implements an agentic DFT subsystem composed of three collaborating agents: CODEX, \cite{chen2021evaluatinglargelanguagemodels} a Form Filler, and a Geometry Reviewer (Figure \ref{fig:lang_to_DFT}). Collectively, they convert natural-language queries
into validated DFT inputs. This framework employs a pure prompting strategy that preserves the flexibility of LLMs while enforcing the precision required for scientific computation. Rather than using rigid templates or parsers, the CODEX agent receives rich contextual information, namely the user's natural language query, detailed ASE~\cite{ase-paper, ase-paper-2} usage patterns for surface science applications, and representative  DFT workflows. At its core, the subsystem relies on intentional context engineering, since the structured inputs and reviewer evaluations allow subsequent agents to detect inconsistencies and correct earlier errors. This enables generalization to new materials, adsorption geometries, and co-adsorption motifs beyond the capabilities of rigid automation. 

Here, we specialize the prompting context to surface–adsorbate systems, which steers the agents toward Cu(100) adatoms and CO adsorption geometries; this reflects the engineered scope of our demonstrations rather than a fundamental limitation of the framework. MASTER can be extended to other structure families, e.g., metal centers embedded in a pre-established N-doped carbon host,~\cite{DAVIS2025145357} by providing the corresponding structural priors in context. Likewise, supplying appropriate examples of input formats enables the simulation agents to target VASP, ~\cite{VASP1, VASP2, VASP3} Quantum Espresso,~\cite{Giannozzi_2009, Giannozzi_2017} ORCA,~\cite{ORCA} or Gaussian.~\cite{gaussian16} Unlike distributed orchestration frameworks \cite{DREAMS2025, MOFGen2024} that coordinate large-scale pipelines, MASTER focuses on scientific reasoning and generates structures through a refinement loop in which stochastic variability is an asset: successive attempts explore alternative configurations, and the reviewer filters them so that only corrected geometries advance. This process links the flexibility of LLMs with first-principles checks and produces reliable results through guided exploration rather than deterministic programming. 

\begin{figure}[!b]
    \centering
    \includegraphics[width=\textwidth]{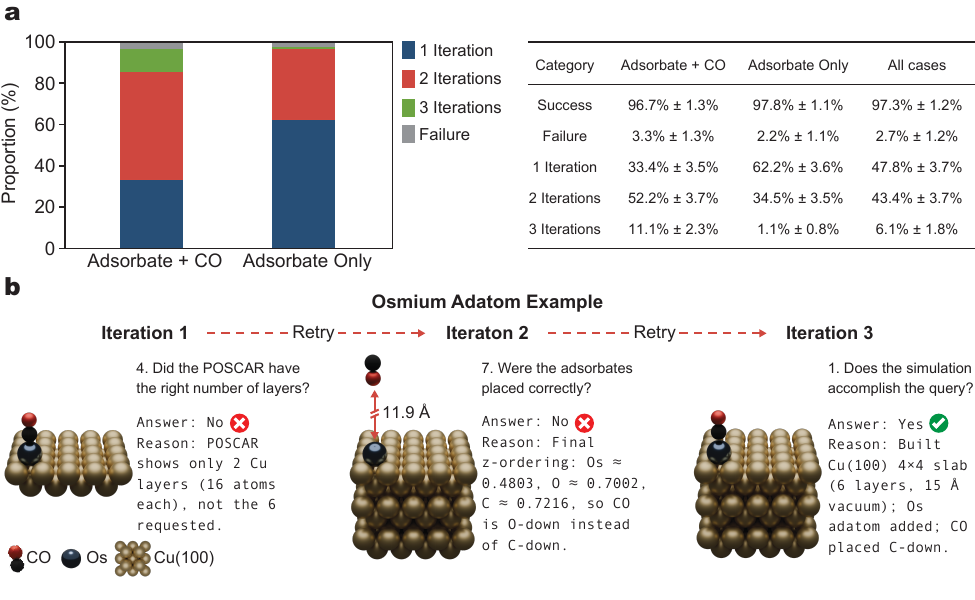}
    \caption{Benchmarking of MASTER simulation agents on transition-metal adatom simulations. \textbf{a}, Performance across 18 representative transition metal-adatom configurations on Cu(100). Each system was generated ten independent times, giving 180 total structure-generation runs that were expert-validated. Bars show the proportion of attempts that converged to a correct geometry in one, two, or three iterations. Summary statistics for success rates are shown in the table on the right. Standard deviations were computed using 5,000 bootstrap resamples. \textbf{b}, Osmium adatom example showing sequential correction across retries. Context built through multimodal evaluation and feedback between agents enables accurate generation and validation of the final geometry.}
    \label{fig:dft_results_example}
\end{figure}

Benchmarking demonstrates that this prompt-based agentic approach achieves a 97.2\% success rate across all test cases after implementing self-revision loops, as shown in Figure \ref{fig:dft_results_example}a (see Methods). Here, success refers to generating a geometry that passes subject matter expert review following the geometry review form (Supplementary Note \ref{note:geo_review}). The subsystem handles both simple adatom placement and more complex adsorption scenarios involving the placement of CO molecules on the desired site. The self-correction loop steadily improves accuracy: 47.8\% of cases succeed on the first iteration, 43.3\% on the second, and 6.1\% on the third. Overall success rates remain high, 97.8\% for adatom-only and 96.7\% for CO-adsorbed systems, with the latter requiring more iterations due to the complexity of molecular orientation (e.g., 11.1\% vs 1.1\% third-iteration convergence). The CO adsorption energies shown in Fig. \ref{fig:our_system_and_materials_problem}b, obtained from fully relaxed DFT calculations, confirm that the constructed geometries are physically meaningful. This strong performance reflects the engineered flow of context between agents, which ensures that information about earlier errors is carried forward and corrected in subsequent attempts.

A representative example of osmium adatom placement on Cu(100) illustrates the iterative refinement process (Figure~\ref{fig:dft_results_example}b). In the first iteration, the generated structure had only two Cu layers instead of the six requested, an error immediately flagged by the Geometry Reviewer. The second iteration corrected the layer count but misplaced the CO molecule with oxygen facing the Os adatom rather than carbon, violating the user's instructions. Only in the third iteration did the system successfully generate the correct structure. This establishes a robust foundation for autonomous DFT execution, demonstrating that LLM agents can reliably translate complex materials specifications into computational workflows without manual intervention.

\subsection*{\textit{LLM Reasoning Strategies for Accelerated Materials Discovery}}

\begin{figure}[!t]
    \centering
    \includegraphics[width=\textwidth]{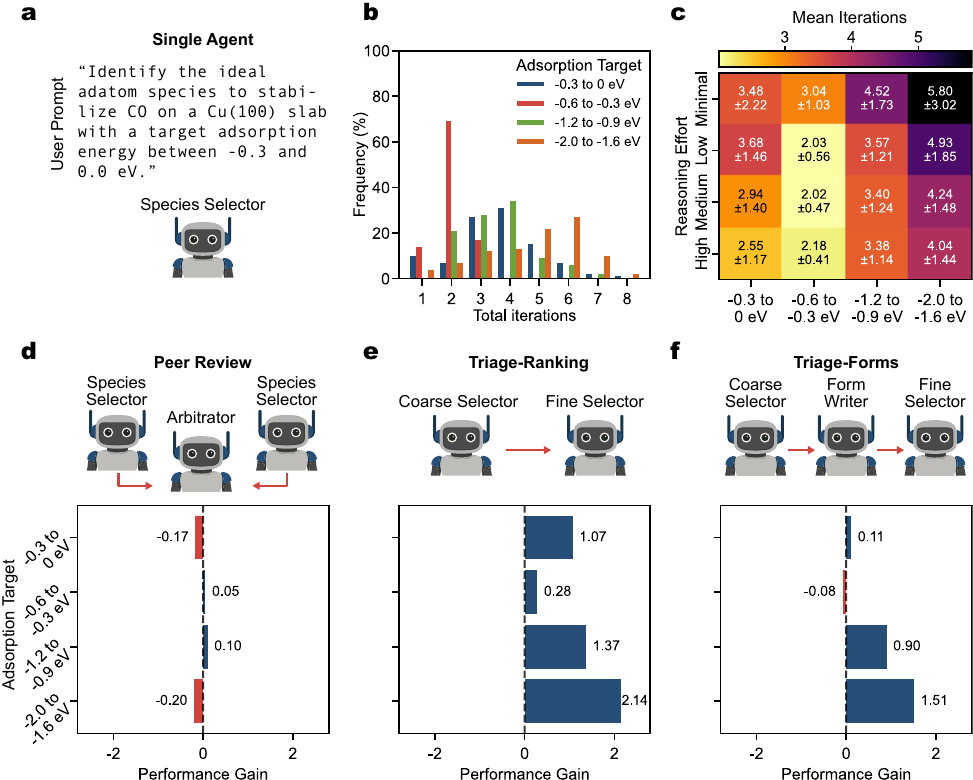}
    \caption{Comparative performance of MASTER's design agents reasoning architectures. \textbf{a}, Schematic of the single agent setup and an example user prompt. \textbf{b}, Frequency distribution of successful iteration counts in the single agent configuration under low reasoning effort (see Methods) \cite{OAI_reasoning_effort}, aggregated over 100 trials for each adsorption energy target, where a success is defined as identifying a material whose adsorption energy falls within the specified target range. \textbf{c}, Heatmap showing the mean iterations to success across adsorption-energy targets and reasoning-effort levels in the single agent configuration. Mean iterations to success heatmaps for the remaining agentic architectures are presented in Figure \ref{fig:Cu100_heatmaps}. \textbf{d}, Performance gain in the peer review multi-agent configuration at high reasoning effort. Performance gain is defined as the improvement in average number of iterations required for success relative to the single agent baseline. \textbf{e}, Performance gain in the triage-ranking configuration at high reasoning effort. \textbf{f}, Performance gain in the triage-forms scenario at high reasoning effort.}
    \label{fig:agentic_reasoning_results} 
\end{figure}

While accurate simulations are essential to the success of the developed agentic approach, the overall efficiency of autonomous discovery depends on how effectively reasoning agents navigate chemical space. Efficient exploration reduces the number of DFT evaluations, whose cost far outweighs that of LLM reasoning. To examine this portion of the workflow, we implemented four agentic reasoning architectures within the MASTER framework. In all configurations, the agents are instructed to apply chemical intuition, drawing on periodic trends, $d$-band theory from their pretraining, and correlations inferred from previous iterations, to propose new material candidates within the same closed loop involving the DFT and reviewer systems described in Fig. \ref{fig:our_system_and_materials_problem}. The strategies differ in how they engineer the context used to generate and refine hypotheses, ranging from single agent reasoning to structured methods of collaboration and hierarchy. The four architectures, illustrated schematically in Fig. \ref{fig:agentic_reasoning_results}, are: 

\begin{description}
    \item [\textbf{Single agent --}] A baseline configuration in which a single LLM autonomously decides the next candidate for evaluation, measuring the capability of one agent to perform self-consistent scientific reasoning without collaboration (Fig. \ref{fig:agentic_reasoning_results}a-c).
    
    \item[\textbf{Peer review --}] A minimal form of collaboration in which two identical but independent agents propose candidates that are reconciled by an arbitrator, testing whether peer oversight improves reliability (Fig. \ref{fig:agentic_reasoning_results}d).
    
    \item[\textbf{Triage-ranking --}] A hierarchical design in which a coarse selector proposes a pool of promising candidates that a fine selector ranks and chooses from, separating exploration from exploitation while retaining chemical diversity (Fig. \ref{fig:agentic_reasoning_results}e).

    \item[\textbf{Triage-forms --}] Similar to triage-ranking but with an additional agent that fills prewritten expert-designed forms for each candidate before the fine selector makes a decision, testing whether guided context improves efficiency over free-form deliberation (Fig.~\ref{fig:agentic_reasoning_results}f).
\end{description}

\begin{figure}[!t]
    \centering
    \includegraphics[width=\textwidth]{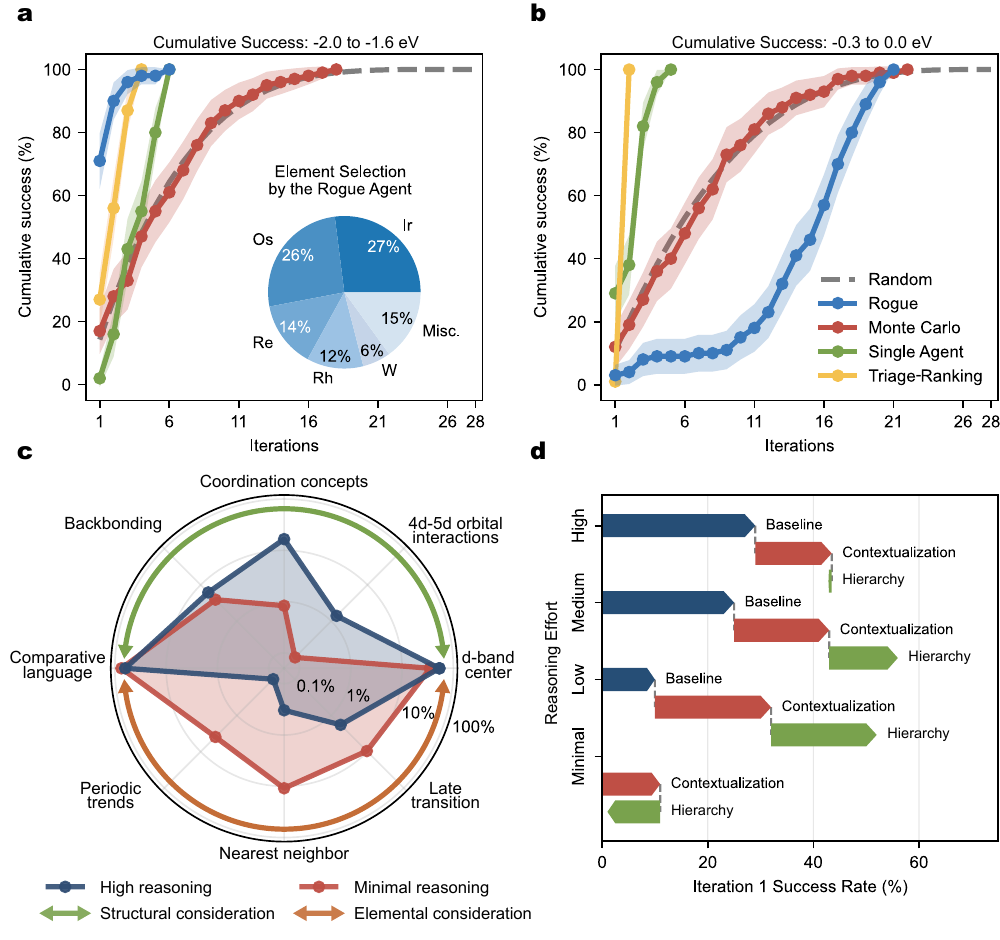}
    \caption{MASTER's cummulative performance, interpretability, and hierarchy effects. \textbf{a}, Success probabilities for the -2 to -1.6 eV target range for Monte Carlo (minimal), rogue (minimal), single (high), and triage-ranking (medium) agents. The label in parenthesis indicates the associated reasoning effort for each architecture. The dashed line shows the theoretical result for trial-and-error selection. ~\cite{Ahlgren2014, Jaynes2003} Inset pie chart shows transition-metal selections made by the rogue agent in the first iteration. Shaded regions denote 95\% confidence intervals using 5,000 bootstrap resamples with the normal approximation (cumulative success probability $\pm$ 1.96 $\times$ bootstrap standard deviation).~\cite{hastie2009elements} \textbf{b}, Cumulative success for the -0.3 to 0 eV target range using the same agents, but with triage-ranking evaluated at minimal reasoning effort. \textbf{c}, Radar plot showing chemical concept categories invoked by the single agent across reasoning-effort levels, aggregated over all iterations and all trials, with categories defined by the keyword sets described in Supplementary Table \ref{tab:keyword_table}. \textbf{d}, Iteration-1 success rates for the -0.3 to 0 eV target range, showing the gains beyond the single agent baseline and associated with instructed enumeration and ranking as well as the multi-agent triage hierarchy.}
    \label{fig:llm_reasoning}
\end{figure}

All reasoning strategies were benchmarked within the above-mentioned transition-metal chemical space, enabling a controlled proof-of-principle testbed for autonomous discovery, as shown in Figs. \ref{fig:agentic_reasoning_results} and \ref{fig:llm_reasoning}.  We use GPT-5~\cite{openai_gpt5_2025} as the base model for all agents. The four target adsorption-energy windows defined in Fig.~\ref{fig:our_system_and_materials_problem}b contain three to five transition metals, ensuring comparable discovery difficulty across binding regimes and allowing differences in performance to be attributed primarily to the reasoning architectures. Equivalent analyses for CO adsorption on M–N–C catalysts are provided in the Supplementary Figures \ref{fig:MNC_heatmaps_SI} - \ref{fig:PC_triage_forms_MNC}. 

Figure \ref{fig:agentic_reasoning_results} summarizes the performance of the four argentic reasoning architectures and their final selected species statistics are presented in Figures \ref{fig:PC_single_agent}-\ref{fig:PC_triage_forms}.  The single agent baseline (Fig. \ref{fig:agentic_reasoning_results}a-c) converges reliably within fewer than ten iterations for all targets, reflecting directed exploration rather than random search. Even alone, the single agent outperforms stochastic baselines (Fig.~\ref{fig:llm_reasoning}a), yielding 100\% cumulative success three times faster than trial-and-error, showing that LLMs already encode physically meaningful priors. The peer review configuration (Fig. \ref{fig:agentic_reasoning_results}d) performs similarly to the single agent, with no discernible performance gain. The species selector agents agreed on the next material in about 60\% of runs while the arbitrator alternated evenly between them otherwise. Their differences were largely stochastic rather than chemically substantive, offering little additional guidance to the arbitrator. This suggests that collective reasoning can only improve performance when agents bring complementary perspectives or distinct priors.

In contrast, the triage-ranking system (Fig. \ref{fig:agentic_reasoning_results}e) achieves the most decisive convergence with an average performance gain as high as 2.14 iterations compared to the single agent baseline. Most runs identify an acceptable adatom within two or three iterations, showing improvement across all target energy windows. Its hierarchical structure balances exploration by the coarse selector with exploitation by the fine selector. By constraining comparison to a small, curated subset, the fine selector receives engineered context while remaining scalable. The triage-forms architecture (Fig. \ref{fig:agentic_reasoning_results}f), which combines a coarse selector, a structured form-filler, and a fine selector, performs slightly below triage-ranking but above the single agent across most energy windows. The best performance was obtained with a form emphasizing relative risk assessment, categorizing each option as a “safe bet”, “moderate risk”, “high risk”, or “unlikely”, identifying the top safe bet when available, and providing a brief rationale (SI Note \ref{note:SM_triage_forms_adatoms}). This design elicits qualitative scientific reasoning, whereas forms that required quantitative predictions, such as $d$-band-center estimates, consistently degraded performance. 

Fig.~\ref{fig:llm_reasoning}a-b presents cumulative success probabilities comparing reasoning architectures with three baselines: a theoretical random sampler, a Monte Carlo agent, and a rogue agent instructed to act randomly but allowed to reason (Methods). Across weak and strong binding windows, the single agent system outperforms the purely stochastic baselines, yielding three fold and eleven fold improvements, respectively. The triage-ranking architecture achieves the steepest success rise, identifying correct candidates within a few iterations and reaching near-unit cumulative probability sooner than any other approach. The rogue agent provides an instructive control: although nominally random, it exhibits a persistent semantic bias toward 5$d$ elements in the early iterations (see pie chart inset). For the strong-binding target (–2.0 to –1.6 eV), this bias fortuitously aligns with the physical trend of increasing CO affinity down the 5$d$ series, yielding apparent outperformance over all other agentic systems. For the weak-binding window (–0.3 to 0 eV), however, the same bias becomes detrimental, steering exploration away from relevant metals and causing success probabilities below even Monte Carlo levels. Cumulative success probabilities for the remaining adsorption energy windows for the Cu(100) case are presented in Figure \ref{fig:cumulative_success_Cu100_SI}. These behaviors reveal that unguided LLM priors can occasionally mimic chemical intuition but remain unreliable without grounding in simulation feedback. Cross-architecture performance differences also reflect a general principle from optimization theory: according to the no-free-lunch theorem~\cite{wolpert1997nfl}, no single search strategy can be optimal across all problem classes. Consistent with this, triage-ranking excels for most Cu(100) targets, whereas other agentic designs perform comparatively or better in the M–N–C catalysts (Figures \ref{fig:MNC_heatmaps_SI} and \ref{fig:cumulative_success_MNC_SI}).

\section*{Discussion}

To understand the behavior of the design agents, we first note that Fig.~\ref{fig:agentic_reasoning_results}c reveals systematic trends with reasoning level. Increasing the GPT-5 reasoning effort generally reduces the mean number of iterations by roughly one, with some targets improving by up to two. For the weak-binding range, for example, minimal-reasoning agents selected Ag first in nearly all runs and reached Au only after several steps, while high-reasoning agents chose Au directly in roughly one-third of cases, reducing the average iterations from $3.5$ to $2.6$. Given the cost of DFT calculations, this reduction represents a meaningful computational gain since each iteration requires two DFT simulations (with and without CO adsorbed). Minimal-reasoning  tends to rely primarily on positional heuristics along the periodic table, whereas high-reasoning agents invoke mechanistic and structural arguments (Fig. \ref{fig:llm_reasoning}c).  For instance, at high reasoning effort references to coordination effects appear nearly an order of magnitude more often and orbital interactions roughly three times as frequent as in minimal-reasoning runs.

We next examine the factors that influence iteration-1 success for the –0.3 to 0 eV adsorption-energy window for the Cu(100) adatom system (Fig.~\ref{fig:llm_reasoning}d). At this stage, the agents have not yet received any DFT results, so success depends solely on how the architecture structures information before feedback. The single agent baseline reflects the model’s ability to propose a plausible candidate from the prompt alone. Adding contextualization, i.e., enumerating all materials and requesting a qualitative ranking, improves accuracy across reasoning levels. At low and medium reasoning effort, the hierarchical multi-agent structure provides an additional gain: the coarse selector’s prescreening and the fine selector’s focused ranking further increase the likelihood of identifying a chemically reasonable first candidate. At high reasoning effort hierarchy offers little additional benefit. Taken together, the results show that the largest gains from hierarchical context engineering arise when reasoning depth is limited or when the candidate set cannot be exhaustively ranked by a single agent. In larger and more heterogeneous spaces, where full enumeration would be impractical, multi-agent hierarchies are therefore expected to play a central role in maintaining high early decision quality.

\begin{figure}[!t]
    \centering
    \includegraphics[width=\textwidth]{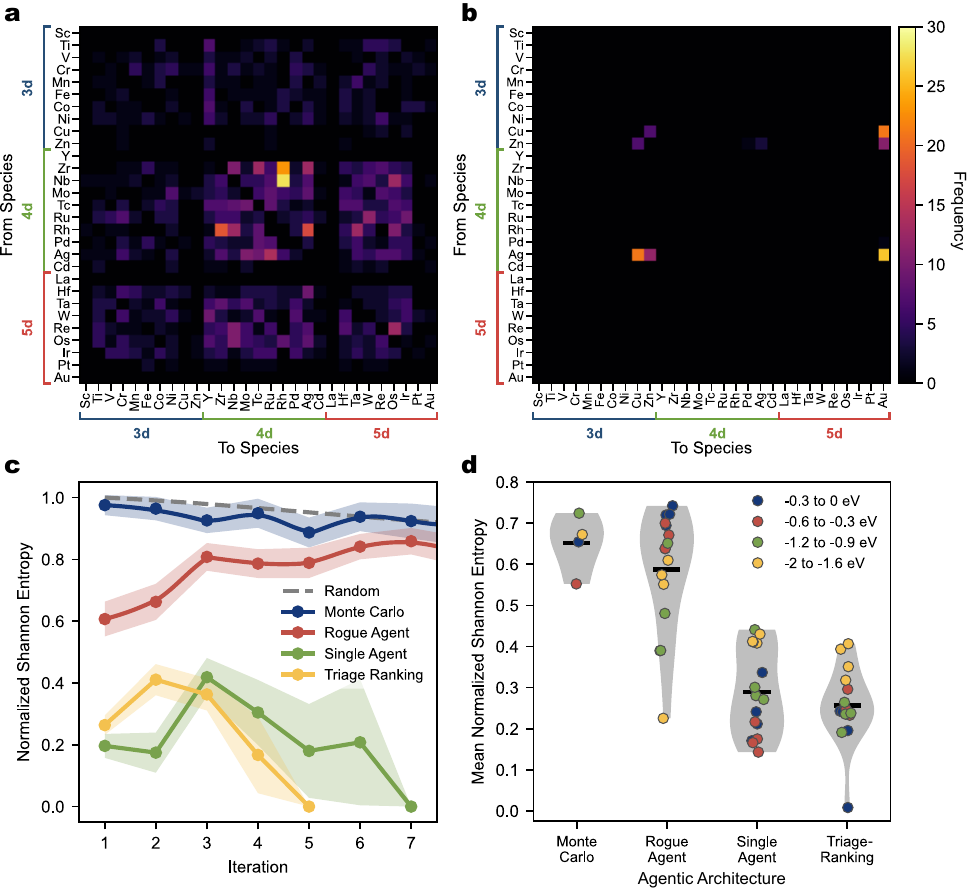}
    \caption{Reasoning trajectories and Shannon entropies across MASTER agent architectures. \textbf{a}, Frequency of transitions between transition metals across consecutive iterations for the -0.3 to 0.0 eV adsorption-energy range on Cu(100) across 100 independent runs for the rogue agent at high reasoning effort. Rows indicate the metal selected in iteration $n$ and columns indicate the metal selected in iteration $n+1$. Colored brackets denote $d$-block periods. \textbf{b}, Species transition heatmap for the triage ranking architecture at high reasoning effort for the same adsorption-energy range. \textbf{c}, Normalized Shannon entropies across iterations for the -0.3 to 0.0 eV adsorption-energy range at medium reasoning effort, with entropy definitions provided in the methods. Shaded regions denote 95\% confidence intervals using 5,000 bootstrap resamples with the normal approximation (Shannon entropy value $\pm$ 1.96 $\times$ bootstrap standard deviation).~\cite{hastie2009elements} \textbf{d}, Mean normalized Shannon entropies for the Monte Carlo agent at minimal reasoning effort and for the rogue agent, single agent, and triage-ranking architectures across all reasoning-effort settings and all adsorption-energy ranges. Equivalent analysis for the M-N-C case is presented in Fig. \ref{fig:reasoning_traj_and_shannon_MNC_SI}.} 
    \label{fig:discussion_heatmaps}
\end{figure}

The reasoning trajectories in Fig.~\ref{fig:discussion_heatmaps} illustrate how agentic hierarchy transforms exploration dynamics in the weak-binding regime. In the rogue agent (Fig.~\ref{fig:discussion_heatmaps}a), an early bias toward 5$d$ metals (Ir, Os, Re, Rh, W) reflects a superficial association between atomic number and adsorption strength. Once these strong-binding early choices fail and the semantically driven prior collapses, the agent explores the space diffusely. By contrast, the triage-ranking agents (Fig.~\ref{fig:discussion_heatmaps}b) exhibits a structured and chemically interpretable transition network concentrated among Ag, Cu, Zn, Ni, and Au, which are elements near the weak-binding window. Frequent Ag $\rightarrow$ Au transitions and recurrent Ag $\rightarrow$ Cu, Ag $\rightarrow$ Zn exchanges indicate that the agents iteratively explore neighboring regions of $d$-band filling. Paths from Ni $\rightarrow$ Au and Cu $\rightarrow$ Au further suggest stepwise correction toward a true weak-binding solution.  
These trajectory patterns rationalize the performance trends observed in Figs.~\ref{fig:agentic_reasoning_results}–\ref{fig:llm_reasoning}. Hierarchical architectures not only accelerate convergence but also reorganize exploration into pathways guided by causal chemical reasoning rather than statistical association.  As the system learns to associate structural and electronic features with adsorption strength, exploration becomes self-correcting, with each iteration reducing uncertainty. In this way, collective reasoning achieves efficiency through progressive information gain rather than exhaustive enumeration.

To quantify the exploration dynamics, we computed the normalized and mean normalized Shannon entropies~\cite{Gray2011Entropy} for each architecture (Fig. \ref{fig:discussion_heatmaps}c-d). Shannon entropy measures how broadly an agentic architecture distributes its selections across the transition metals in each iteration, with lower values indicating a more focused and information-efficient search. \cite{original_Shannon, 50_years_shannon} The rogue agent and Monte Carlo baselines maintain persistently high entropy, consistent with their erratic exploration. By contrast, both single agent and triage-ranking architectures exhibit pronounced entropy contraction, with the latter  achieving the lowest mean entropy overall. These results show that the hierarchical architectures outperform the single agent baseline because the context they propagate systematically provides information advantage that guides the search toward the correct region of the search space in fewer iterations.  

In the present MASTER implementation, the design agents have no \emph{a priori} knowledge of the absolute adsorption-energy scale or level of theory used in our DFT calculations. They must instead infer these scales on the fly from the sequence of simulation outcomes and the acceptance criteria, learning which regimes correspond to weak, intermediate, or strong binding. In small, fully enumerable spaces such as our adatom benchmark, this implicit calibration is sufficient. In larger or less well-characterized domains, however, an additional retrieval-augmented agent could supply prior grounding by querying literature or materials databases, improving robustness and accelerating convergence when simulations are costly or the underlying energy landscape is complex. Similarly, structured, form-based triage is  likely to become more valuable in such regimes, where standardized prompts can stabilize reasoning, enforce consistent comparison criteria, and preserve interpretability across many interacting agents. 

Altogether, our findings show that structured agentic collaboration transforms large language models from procedural tools into adaptive scientific reasoners. Within MASTER, autonomy arises from interaction: agents that deliberate, incorporate feedback, and refine shared hypotheses guide exploration with increasing mechanistic consistency. By linking language, simulation, and theory into a unified workflow, MASTER enables efficient, self-correcting discovery. Across the CO-adsorption problems studied here, this combination of hierarchical reasoning and autonomous simulation reduces the number of required atomistic calculations by up to 90\% relative to trial-and-error while preserving first-principles accuracy. Extending such architectures beyond materials science could enable general-purpose scientific agents capable of autonomous hypothesis formation and reasoning across the physical and life sciences.

\section*{Methods}

\subsection*{\textit{Atomistic Simulations using Density Functional Theory}}

All DFT calculations for the transition metal-adatom case were performed using the Vienna Ab initio Simulation Package~\cite{VASP1, VASP2, VASP3} (VASP, version 6.4.2) within the projector augmented-wave (PAW) formalism. We employed the revised Perdew–Burke–Ernzerhof (RPBE) functional \cite{PhysRevB.59.7413} within the generalized gradient approximation (GGA) to describe exchange–correlation contribution to the system Hamiltonian. A plane-wave energy cutoff of 580 eV was used, and all calculations were spin-polarized. We modeled the Cu(100) surface as a six-layer, 4 $\times$ 4 periodic slab containing 96 Cu atoms, separated by a 15 Å vacuum region. A single transition-metal adatom was positioned in the fourfold hollow site of the surface, and CO was adsorbed atop the adatom.

Brillouin-zone integrations were performed using a 4 $\times$ 4 $\times$ 1 Monkhorst–Pack $k$-point mesh, which was verified to yield converged adsorption energies within 0.01 eV. All structures were optimized until the forces on unconstrained atoms were below $0.02\ \text{eV}\ \text{Å}^{-1}$ and electronic convergence was achieved to within $10^{-6}$ eV. The bottom two Cu layers were held fixed to their bulk positions, while all other atoms were allowed to relax. Adsorption energies were determined from total electronic energies of the fully relaxed structures according to Eq. (\ref{eq:1}):
\begin{equation}
E_{\text{ads}} = E_{\text{CO/M/Cu(100)}} - E_{\text{M/Cu(100)}} - E_{\text{CO(g)}},
\label{eq:1}
\end{equation}
where $E_{\text{CO/M/Cu(100)}}$, $E_{\text{M/Cu(100)}}$, and $E_{\text{CO(g)}}$ are the total electronic energies of the CO-adsorbed system, the M-decorated Cu(100) slab, and the isolated CO molecule, respectively (see Table \ref{tab:co_adsorption_cu100}). All reported adsorption energies correspond to electronic energies at 0~K, without zero-point or entropic corrections which are left to future work but should not change the qualitative nature of the findings. Negative values of $E_{\text{ads}}$ indicate exothermic adsorption.

CO adsorption energies on M-$\mathrm{N_4C_{10}}$ catalysts are computed using DFT as previously reported.~\cite{DAVIS2025145357} An initial Fe-$\mathrm{N_4C_{10}}$ structure with 66 total carbon atoms is first relaxed, then starting structures for all transition metals are generated using ASE \cite{ase-paper} by replacing Fe with a given transition metal. Calculations are carried out with VASP using the RPBE functional and default PBE projector augmented wave-pseudopotentials \cite{PhysRevB.50.17953, PhysRevB.59.1758} and managed with the pyiron workflow framework. \cite{pyiron-paper} A cell size of 14.78 Å $\times$ 12.80 Å is used for all surface calculations with a 20 Å vacuum normal to the surface. A 4 $\times$ 4 $\times$ 1 Monkhorst-Pack $k$-point mesh is employed with dipole corrections applied normal to the surface. Spin polarization is turned on for all calculations. The plane-wave basis cutoff is set to 600 eV, and a Fermi-Dirac smearing width of 0.0259 is used. During structural relaxation, only atomic positions are allowed to relax, while the cell volume and shape remain fixed. Geometries are converged to a threshold of $<10^{-5}$ eV change in energy between sequential steps. The gas-phase energy of CO is computed by placing the molecule in the center of the same size unit cell as the M-$\mathrm{N_4C_{10}}$ structures and allowed the atoms to relax.

For structural relaxation of the surfaces, the planar initial structure and a structure with the transition metal center displaced 0.6 Å out of plane are both relaxed. This is done to avoid trapping in high-energy meta-stable configurations; the lower-energy optimized is used as the reference structure for subsequent CO adsorption calculations. CO-adsorbed structures are generated by placing CO above the transition metal center in three initial configurations: with the carbon atom bound to the surface and the oxygen atom in line with vector normal to the surface, with the oxygen atom bound to the surface and the carbon atom in line with vector normal to the surface, and a bidentate configuration with the C-O bond positioned directly above the transition metal and oriented parallel to the surface. The adsorbate structure which yields the lowest overall energy is then used for computing adsorption energy similarly to Eq. \ref{eq:1} but replacing M/Cu(100) $\rightarrow$ M-N-C.

\subsection*{\textit{LLM Framework for Density Functional Theory Simulations}}

The atomic position generation component of MASTER uses a three-agent workflow built on OpenAI Agents SDK (version 0.0.18).~\cite{openai_agents_sdk_2025} The Geometry Generator agent receives a natural language structure request and constructs a prompt containing the user query plus a JSON knowledge base with twelve ASE construction tips covering site placement, molecular orientation, and covalent radii for common surface atoms and adsorbates (SI Note \ref{note:geo_tips}). This prompt is passed to Codex (version 0.57.0)~\cite{chen2021evaluatinglargelanguagemodels} via command-line interface as a subprocess, which returns a Python script using ASE library functions. The script executes in an isolated temporary directory to produce a VASP POSCAR file. The Geometry Generator agent then creates three orthogonal structure visualizations (top, side, and profile views) and transfers control to the Form Filler agent, which accesses outputs through shared filesystem directories.

The Form Filler agent analyzes the POSCAR file to verify atomic composition and layer count, examines the three visualization images, and completes an eight-question binary assessment (SI Note \ref{note:geo_review}) evaluating composition, layer constraints, site placement, orientation, and vertical spacing. For the current surface adsorption application, assessment questions include domain-specific criteria such as ``was the adsorbate placed in the right place" (evaluating hollow, bridge, or on-top site occupancy) and ``were the adsorbates placed in the right orientation" (verifying molecular geometry such as C-down vs O-down for CO). The form template and construction tips are modular components that can be modified for other simulation domains by replacing the assessment criteria and construction guidelines while preserving the three-agent workflow architecture. 

The completed form transfers to the Reviewer agent, which inspects the images and POSCAR file for consistency with the form assessment and makes the final acceptance decision. Structures pass only if all eight questions receive affirmative responses and required files exist. Rejection triggers written feedback specifying identified deficiencies and returns control to the generator agent with incremented version numbering. The generator agent incorporates this feedback into revised Codex prompts for iterative refinement, supporting up to five cycles. Rejected structures archive to version-controlled subdirectories preserving the complete revision history. All agent decisions, Codex prompts, and generated scripts log to structured markdown files for reproducibility.

\subsection*{\textit{Benchmarking Protocol for Agentic LLM Reasoning}}

We benchmarked four reasoning strategies in MASTER using the OpenAI Agents SDK with GPT-5 models.~\cite{openai_gpt5_2025} All agents operated at low verbosity, and the reasoning effort parameter, as implemented by OpenAI, was swept across minimal, low, medium, and high settings. \cite{OAI_reasoning_effort} The search space for both the adatom and M-N-C test cases comprised the 28 transition metals from Sc to Au. Targets were specified as numerical adsorption-energy bands (e.g., -0.6 eV to -0.3 eV) without tolerance, and each run terminated when the reviewer confirmed that the measured energy laid within the specified band. Each strategy was executed as independent batches spanning four adsorption-energy windows (SI Note \ref{note:queries}). For every window and reasoning-effort level, one hundred runs were recorded. Within each run, the selector proposed an untested element from the Sc–Au set, the evaluator returned a fixed ground-truth adsorption energy from DFT computations, and the reviewer determined whether the band criterion had been met. Invalid or duplicate proposals were rejected and re-prompted up to a total of 20 maximum retries. The trial is considered a failure if the system reaches the maximum number of retries. In all tests we preformed in this paper, we never observed a failed trial. The prompts and system messages used for all agentic architectures are presented in SI Notes \ref{note:SM_single_agent_adatoms}-\ref{note:SM_triage_forms_adatoms} for the transition metal adamon on Cu(100) case and SI Notes \ref{note:SM_single_agent_MNC}-\ref{note:SM_triage_forms_MNC} for the M-N-C case. 

To establish a theoretical baseline for comparison, we computed the cumulative probability of success for random sampling without replacement as~\cite{Ahlgren2014, Jaynes2003} 

\begin{equation}
P_{\text{success}}(i) = 1 - \frac{\binom{N-K}{i}}{\binom{N}{i}}, 
\label{eq:probability_random}
\end{equation}
where $N=28$  is the total number of transition metals candidates, $K$ is the number of correct materials within the target energy band, $i$ is the iteration index, and $\binom{n}{k}$ denotes the binomial coefficient.

In addition to the reasoning-based strategies, we also implemented two control agents to assess the impact of strategic reasoning versus pure randomness. The Monte Carlo agent employs a deterministic random number generator tool that selects a uniform random index from 1 to 28, mapping each index to the corresponding transition metal in the Sc–Au series (SI Note \ref{note:SM_monte_carlo}). The agent is instructed to call this tool without applying any materials science reasoning, providing a computationally controlled random baseline. The rogue agent, by contrast, is an LLM-based selector instructed to perform completely random selection with no strategic reasoning, bias, or optimization (SI Note \ref{note:SM_rogue_agent}). The rogue agent is explicitly prohibited in prompting from using materials science concepts (periodic trends, $d$-band theory, electronegativity) and is directed to select candidates as if rolling dice, establishing an LLM-based random baseline that tests whether the model can suppress its reasoning priors when instructed to do so.

\subsection*{\textit{LLM Chemical Concept Usage Analysis}}

To quantify how GPT-5 reasoning effort influences performance, we analyzed the chemical concepts invoked by the single agent system across all chemical targets (Fig. \ref{fig:our_system_and_materials_problem}b) for minimal and high reasoning effort. This analysis enables us to determine how the model’s conceptual grounding shifts with reasoning effort and directly supports the comparison shown in Figure \ref{fig:llm_reasoning}c. We defined eight categories representing distinct chemical concepts, with each category encompassing multiple keyword variants to capture linguistic variations (Table \ref{tab:keyword_table}). These categories can be broadly grouped into structural concepts (coordination effects, backbonding, 4$d$/5$d$ orbital interactions) and elemental concepts (periodic trends, nearest neighbor, and late transition metal classifications). The keyword matching procedure performs case-insensitive substring searches within each reasoning statement, counting a statement as positive for a category if any variant appears. Each statement contributes at most one count per category, preventing double-counting when multiple variants of the same concept appear in a single statement.

Frequency calculations proceed by dividing the number of statements containing each category's keywords by the total number of statements analyzed, expressed as percentages. For minimal reasoning, we analyzed 1,684 statements across all energy windows and runs; for high reasoning, 1,215 statements. The resulting percentages represent the fraction of selection decisions that invoked each chemical concept category.

\subsection*{\textit{Shannon Entropy}}

To analyze the heterogeneity of species selected across iterations, we computed the normalized Shannon entropy for each iteration (\ref{eq:norm_shannon}).\cite{original_Shannon, 50_years_shannon} We first construct the probability distribution $p_j^{(i)}$ based on the frequency with which each species $j$ was selected across all active runs at iteration $i$. The Shannon entropy $H_i$ is then computed as~\cite{Gray2011Entropy} 
\begin{equation}
H_i = -\sum_{j=1}^{N} p_j^{(i)} \log(p_j^{(i)}),
\label{eq:shannon}
\end{equation}
where $N=28$ is the total number of candidate species. To normalize the Shannon entropy to the range [0,1], we divide by the maximum possible entropy $\log(N)$:
\begin{equation}
H_{\text{norm}}^{(i)} = \frac{H_i}{\log{N}}.
\label{eq:norm_shannon}
\end{equation}

This normalized entropy equals 1 when all species are equally likely to have been chosen (maximum heterogeneity) and approaches 0 as selection becomes concentrated on fewer species (lower heterogeneity). To characterize the overall exploration behavior of each agentic architecture across an entire run, we computed the mean normalized Shannon entropy $\bar{H}_{\text{norm}}$ over $T$ iterations:
\begin{equation}
\bar{H}_{\text{norm}} = \frac{1}{T} \sum_{i=1}^{T} H_{\text{norm}}^{(i)}.
\label{eq:mean_normalized_shannon}
\end{equation}

This metric represents the average normalized Shannon entropy per iteration, providing a single value that captures the typical heterogeneity level maintained throughout the selection process. 

\section*{Declarations}

\subsection*{Acknowledgments}
This research was supported by the Institute for Materials Science of Los Alamos National Laboratory. Research presented in this article was supported by the Laboratory Directed Research and Development program of Los Alamos National Laboratory under project number 20230065DR. This research used resources provided by the Los Alamos National Laboratory Institutional Computing Program, which is supported by the U.S. Department of Energy National Nuclear Security Administration under Contract No. 89233218CNA000001. This material is also based upon work supported by the U.S. Department of Energy, Office of Critical Minerals and
Energy Innovation (CMEI), specifically the Hydrogen and Fuel Cell Technologies Office (HFTO) under contract ELY-BIL003. Samuel Rothfarb received a UConn’s Pratt \& Whitney Institute for Advanced Systems Engineering Graduate Fellowship which enabled Samuel to contribute to this work.

\subsection*{Conflict of Interest}

The authors declare no conflicts of interest.

\subsection*{Code Availability}

The code supporting this work is available from the corresponding authors upon reasonable request. 

\subsection*{Data Availability}

The data supporting this work is provided in the Supplementary Information. 

\subsection*{Author Contribution}

S.R., B.L., E.F.H., and W.K.K. conceptualized the project. S.R. developed density functional theory calculations for adatoms on Cu(100) under guidance from E.F.H. and the reasoning strategies under guidance from W.K.K. M.D. and I.M performed benchmark DFT calculations for the M-N-C systems. S.R. and W.K.K. wrote the paper and received feedback from all authors, who reviewed and approved its final version. B.L., E.F.H, and W.K.K, supervised the project execution.  

\printbibliography[title={References}]

\newpage

\setcounter{figure}{0}
\renewcommand{\thefigure}{S\arabic{figure}}

\setcounter{table}{0}
\renewcommand{\thetable}{S\arabic{table}}

\setcounter{page}{1}
\renewcommand{\thepage}{S-\arabic{page}}

\newcounter{suppnote}
\newcommand{\suppnote}[1]{%
  \refstepcounter{suppnote}%
  \subsection*{Supplementary Note \thesuppnote. #1}}

\begin{titlepage}
\thispagestyle{plain}

\centering

\vspace*{4ex}

{\sffamily\bfseries\LARGE
Supplementary Information for \\ Hierarchical Multi-agent Large Language Model Reasoning for Autonomous Functional Materials Discovery\\[3ex]
}

{\large \sffamily
Samuel Rothfarb$^{\dagger,\ddagger}$ \orcidlink{0009-0004-6696-9495}, 
Megan C. Davis$^\ddagger$ \orcidlink{0000-0002-4038-8615}, 
Ivana Matanovic$^\ddagger$ \orcidlink{0000-0002-9191-8620},\\[1ex]
Baikun Li$^\dagger$* \orcidlink{0000-0002-5623-5912}, 
Edward F. Holby$^\ddagger$* \orcidlink{0000-0001-8419-6298}, and 
Wilton J.M. Kort-Kamp$^\ddagger$* \orcidlink{0000-0002-0679-6690}\\[2ex]
}

{\normalsize \textit{
$^\dagger$School of Civil \& Environmental Engineering, University of Connecticut, Storrs, Connecticut 06269, United States.\\
$^\ddagger$Theoretical Division, Los Alamos National Laboratory, Los Alamos, New Mexico 87545, United States.\\[2ex]
}}

{\sffamily
*Corresponding authors: 
\href{mailto:baikun.li@uconn.edu}{baikun.li@uconn.edu}, 
\href{mailto:holby@lanl.gov}{holby@lanl.gov}, 
\href{mailto:kortkamp@lanl.gov}{kortkamp@lanl.gov}\\
}

\vfill
\end{titlepage}

\begin{figure}[H]
    \centering
    \includegraphics[width=\textwidth]{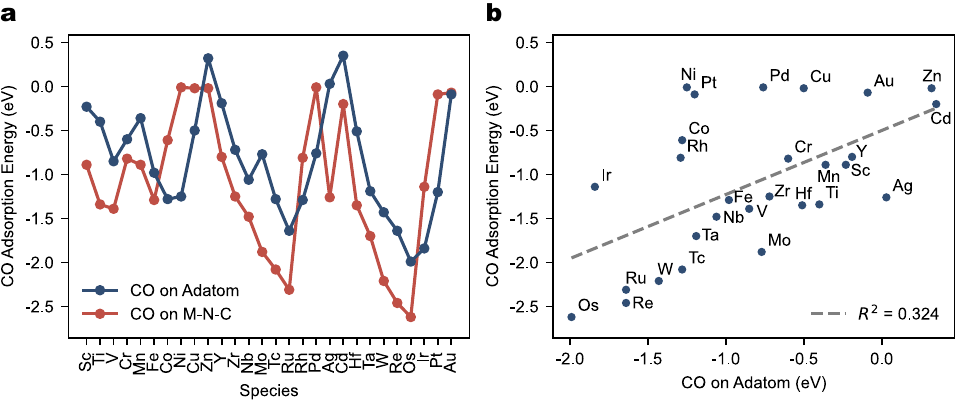}
    \caption{\textbf{a}, CO adsorption energies across metal species for  transition metal adatom on Cu(100) sites and M-N-C catalysts. \textbf{b}, Correlation between CO adsorption energies for the two cases considered here, with linear fit ($R^2$ shown). }
    \label{fig:adatom_vs_mnc}
\end{figure}


\begin{figure}[H]
    \centering
    \includegraphics[width=\textwidth]{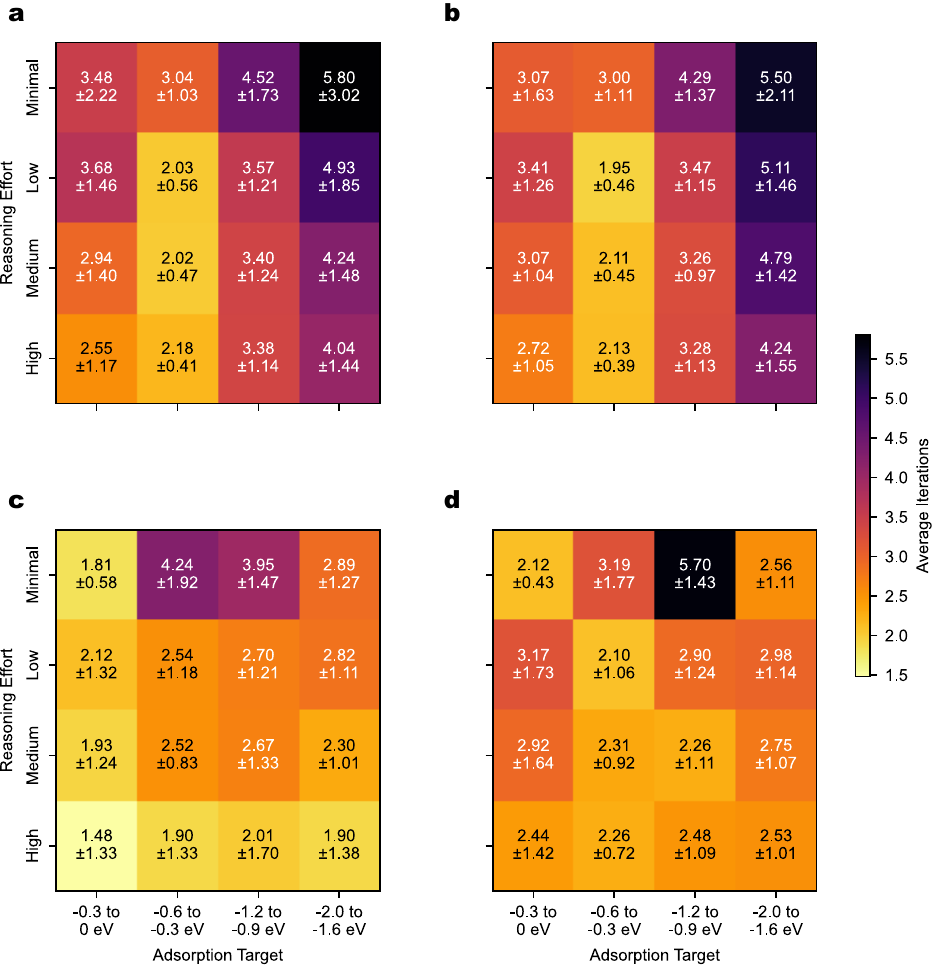}
    \caption{Heatmaps showing mean iterations to success for the Cu(100) test case for the \textbf{a}, single agent, \textbf{b}, peer review, \textbf{c}, triage-ranking, and \textbf{d}, triage-forms configurations.}
    \label{fig:Cu100_heatmaps}
\end{figure}

\begin{figure}[H]
    \centering
    \includegraphics[width=\textwidth]{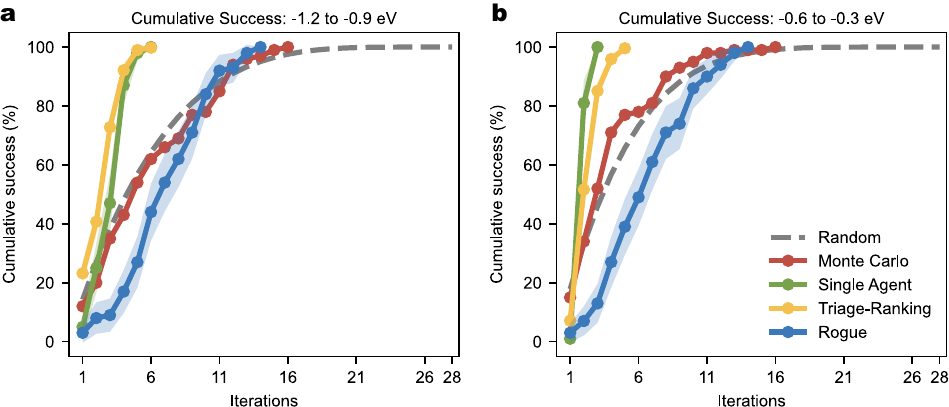}
    \caption{MASTER's cummulative performance on Cu(100) cases. \textbf{a}, Success probabilities for the -1.2 to -0.9 eV target range for Monte Carlo (minimal), rogue (minimal), single (high), and triage-ranking (medium) agents. The label in parenthesis indicates the associated reasoning effort for each architecture. The dashed line shows the theoretical result for trial-and-error selection. ~\cite{Ahlgren2014, Jaynes2003} Inset pie chart shows transition-metal selections made by the rogue agent in the first iteration. Shaded regions denote 95\% confidence intervals using 5,000 bootstrap resamples with the normal approximation (cumulative success probability $\pm$ 1.96 $\times$ bootstrap standard deviation).~\cite{hastie2009elements} \textbf{b}, Cumulative success for the -0.6 to -0.3 eV target range using the same agent architectures and reasoning effort.}
    \label{fig:cumulative_success_Cu100_SI}
\end{figure}

\begin{figure}[H]
    \centering
    \includegraphics[width=\textwidth]{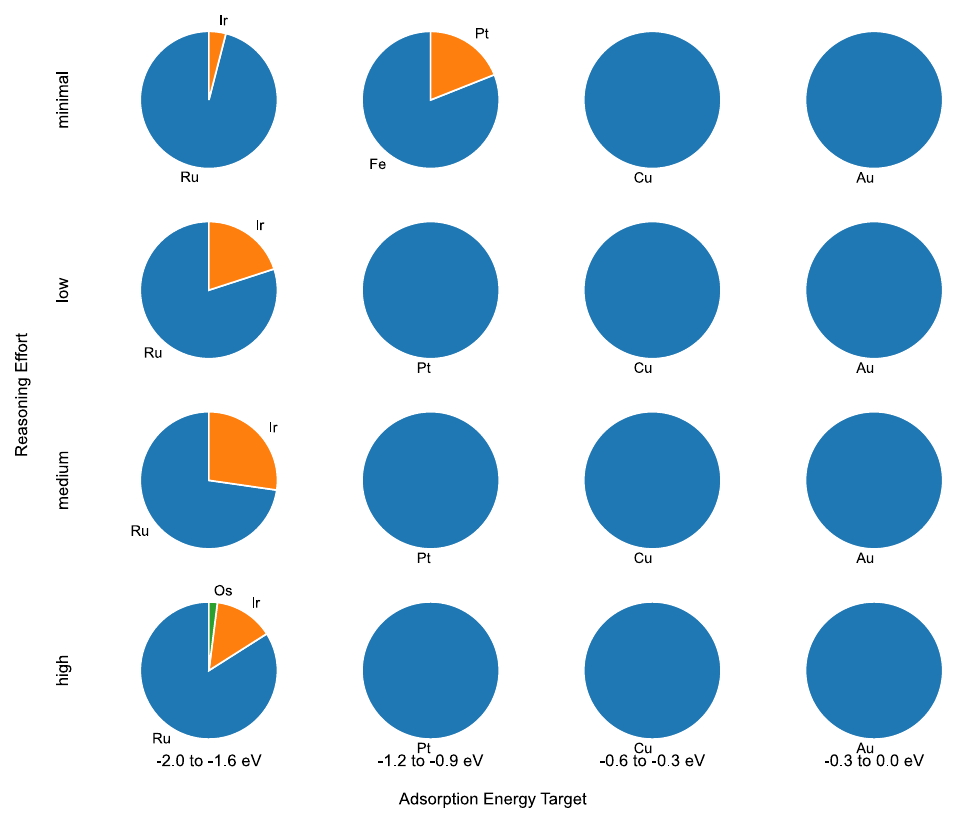}
    \caption{Breakdown of the final species chosen by the single agent architecture by reasoning effort and adsorption energy target range for CO adsorption on adatoms.}
    \label{fig:PC_single_agent}
\end{figure}

\begin{figure}[H]
    \centering
    \includegraphics[width=\textwidth]{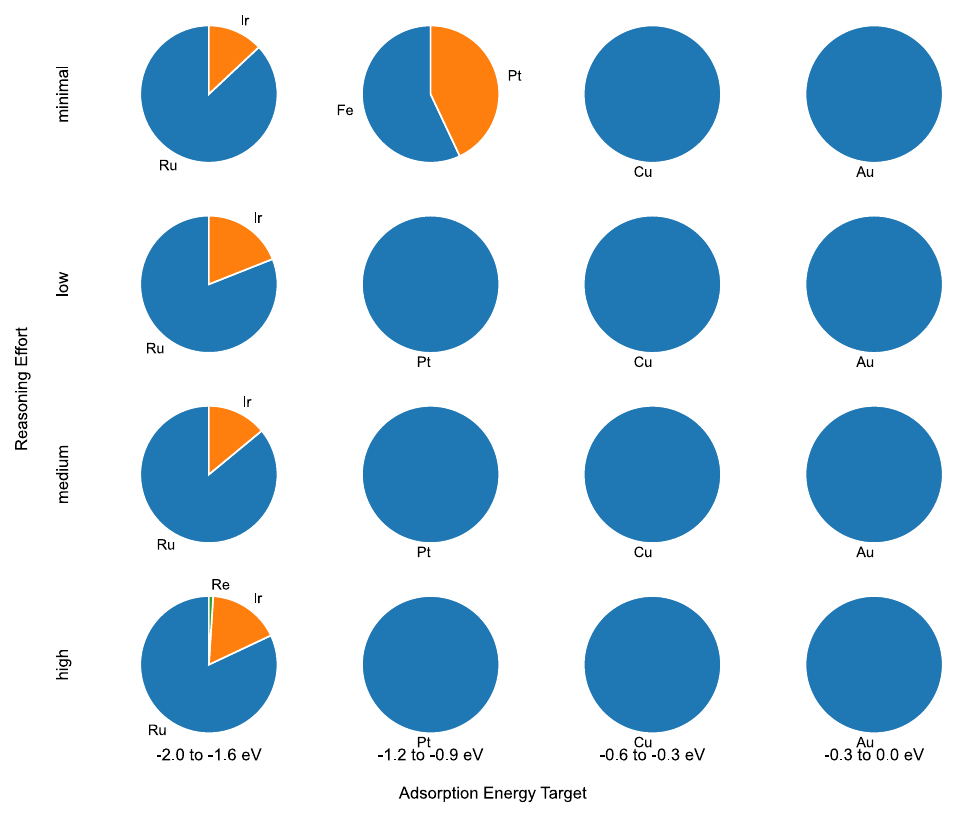}
    \caption{Breakdown of the final  species chosen by the peer review architecture by reasoning effort and adsorption energy target range for CO adsorption on adatoms.}
    \label{fig:PC_peer_review}
\end{figure}

\begin{figure}[H]
    \centering
    \includegraphics[width=\textwidth]{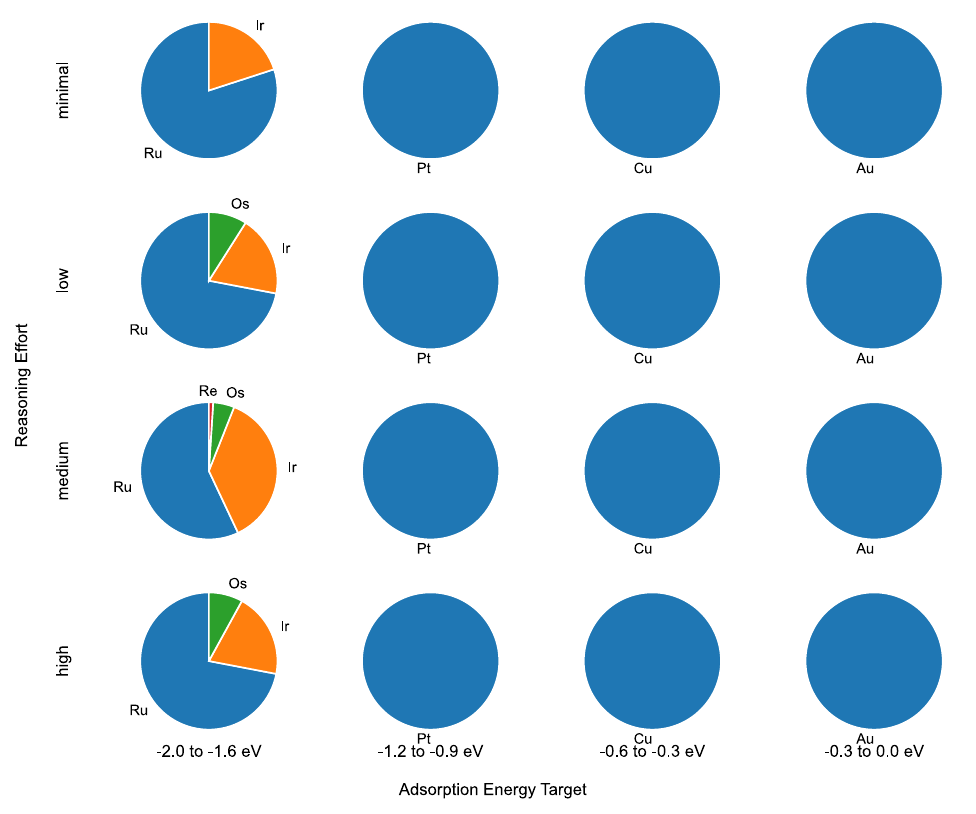}
    \caption{Breakdown of the final species chosen by the triage-ranking architecture by reasoning effort and adsorption energy target range for CO adsorption on adatoms.}
    \label{fig:PC_triage_ranking}
\end{figure}

\begin{figure}[H]
    \centering
    \includegraphics[width=\textwidth]{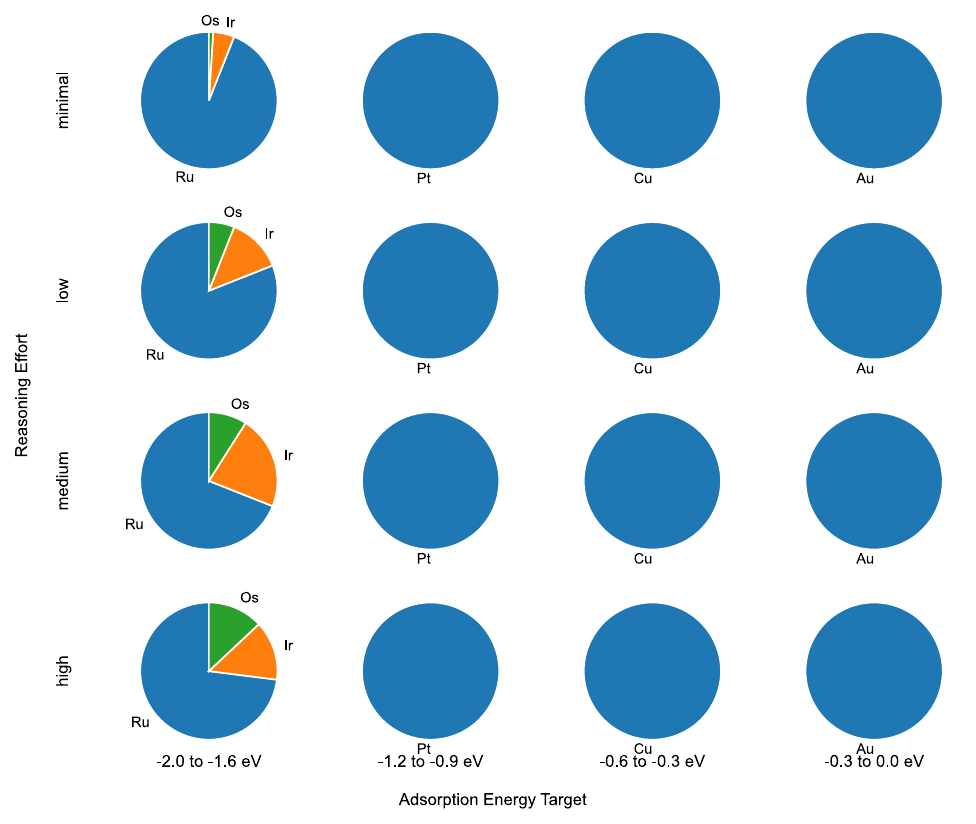}
    \caption{Breakdown of the final species chosen by the triage-forms architecture by reasoning effort and adsorption energy target range for CO adsorption on adatoms.}
    \label{fig:PC_triage_forms}
\end{figure}


\begin{figure}[H]
    \centering
    \includegraphics[width=\textwidth]{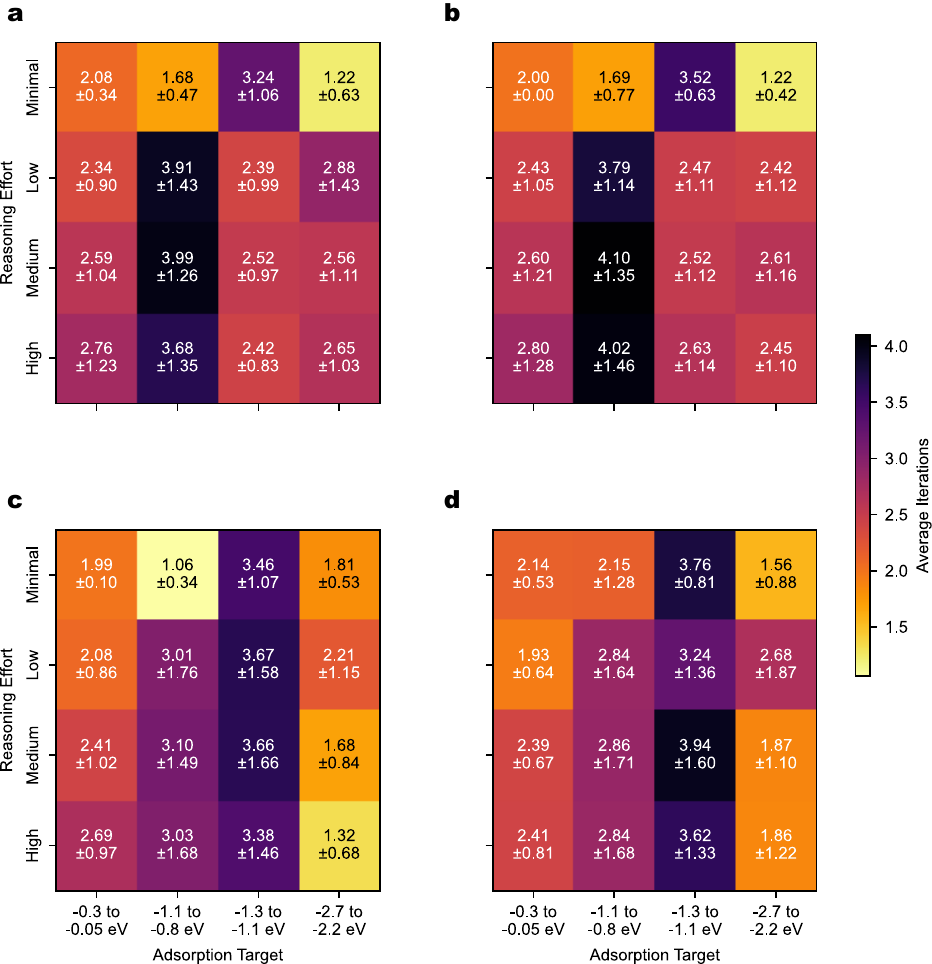}
    \caption{Heatmaps showing mean iterations to success for the M-N-C test case for the \textbf{a}, single agent, \textbf{b}, peer review, \textbf{c}, triage-ranking, and \textbf{d}, triage-forms configurations.}
    \label{fig:MNC_heatmaps_SI}
\end{figure}

\begin{figure}[H]
    \centering
    \includegraphics[width=\textwidth]{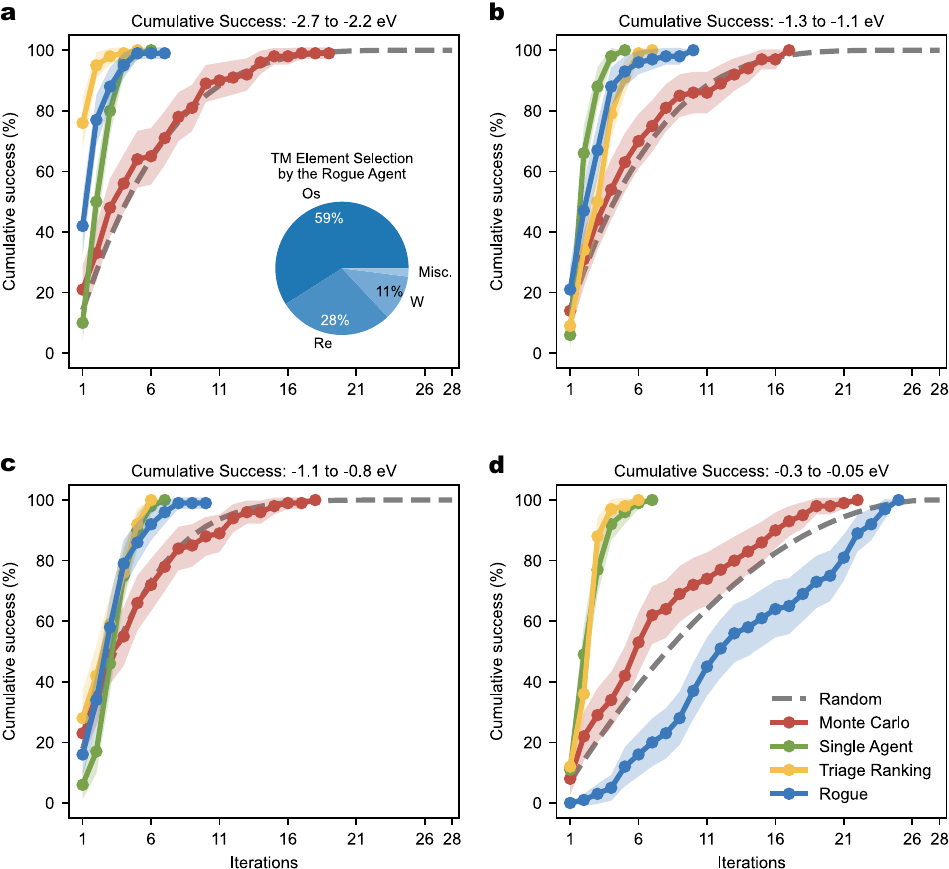}
    \caption{MASTER's cummulative performance on M-N-C cases. \textbf{a}, Success probabilities for the -2.7 to -2.2 eV target range for Monte Carlo (minimal), rogue (minimal), single (high), and triage-ranking (high) agents. The label in parenthesis indicates the associated reasoning effort for each architecture. The dashed line shows the theoretical result for trial-and-error selection. ~\cite{Ahlgren2014, Jaynes2003} Inset pie chart shows transition-metal selections made by the rogue agent in the first iteration. Shaded regions denote 95\% confidence intervals using 5,000 bootstrap resamples with the normal approximation (cumulative success probability $\pm$ 1.96 $\times$ bootstrap standard deviation).~\cite{hastie2009elements} \textbf{b}, Cumulative success for the -1.3 to -1.1 eV target range using the same agent architectures and reasoning effort. \textbf{c}, Cumulative success for the -1.1 to -0.8 eV target range using the same agent architectures and reasoning effort. \textbf{d}, Cumulative success for the -0.3 to -0.05 eV target range using the same agent architectures and reasoning effort.}
    \label{fig:cumulative_success_MNC_SI}
\end{figure}

\begin{figure}[H]
    \centering
    \includegraphics[width=\textwidth]{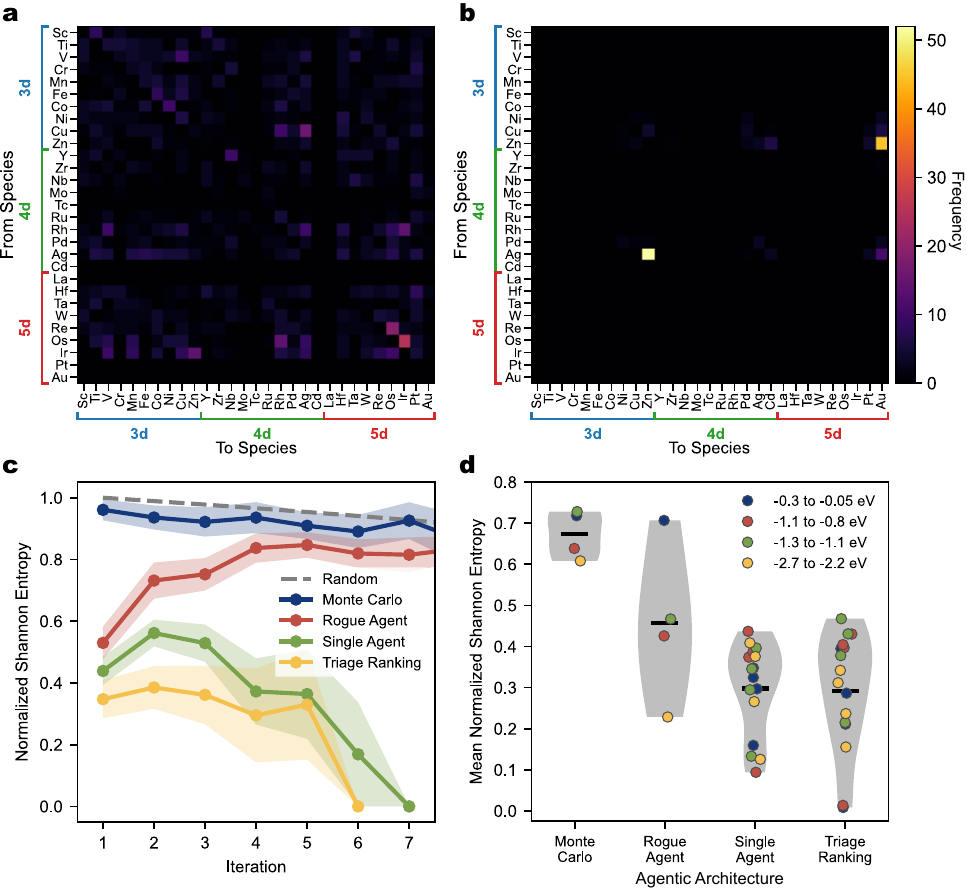}
    \caption{Reasoning trajectories and Shannon entropies across MASTER agent architectures. \textbf{a}, Frequency of transitions between transition metals across consecutive iterations for the -0.3 to -0.05 eV adsorption-energy range on M-N-C across 100 independent runs for the rogue agent at minimal reasoning effort. Rows indicate the metal selected in iteration $n$ and columns indicate the metal selected in iteration $n+1$. Colored brackets denote $d$-block periods. \textbf{b}, Species transition heatmap for the triage ranking architecture at high reasoning effort for the same adsorption-energy range. \textbf{c}, Normalized Shannon entropies across iterations for the -0.3 to -0.05 eV adsorption-energy range with Monte Carlo and rogue agent architectures at minimal reasoning effort and single agent and triage-ranking architectures at high reasoning effort, with entropy definitions provided in the Methods. Shaded regions denote 95\% confidence intervals using 5,000 bootstrap resamples with the normal approximation (Shannon entropy value $\pm$ 1.96 $\times$ bootstrap standard deviation).~\cite{hastie2009elements} \textbf{d}, Mean normalized Shannon entropies for the Monte Carlo and rogue agent architectures at minimal reasoning effort and for single agent, and triage-ranking architectures across all reasoning-effort settings and all adsorption-energy ranges.}
    \label{fig:reasoning_traj_and_shannon_MNC_SI}
\end{figure}

\begin{figure}[H]
    \centering
    \includegraphics[width=\textwidth]{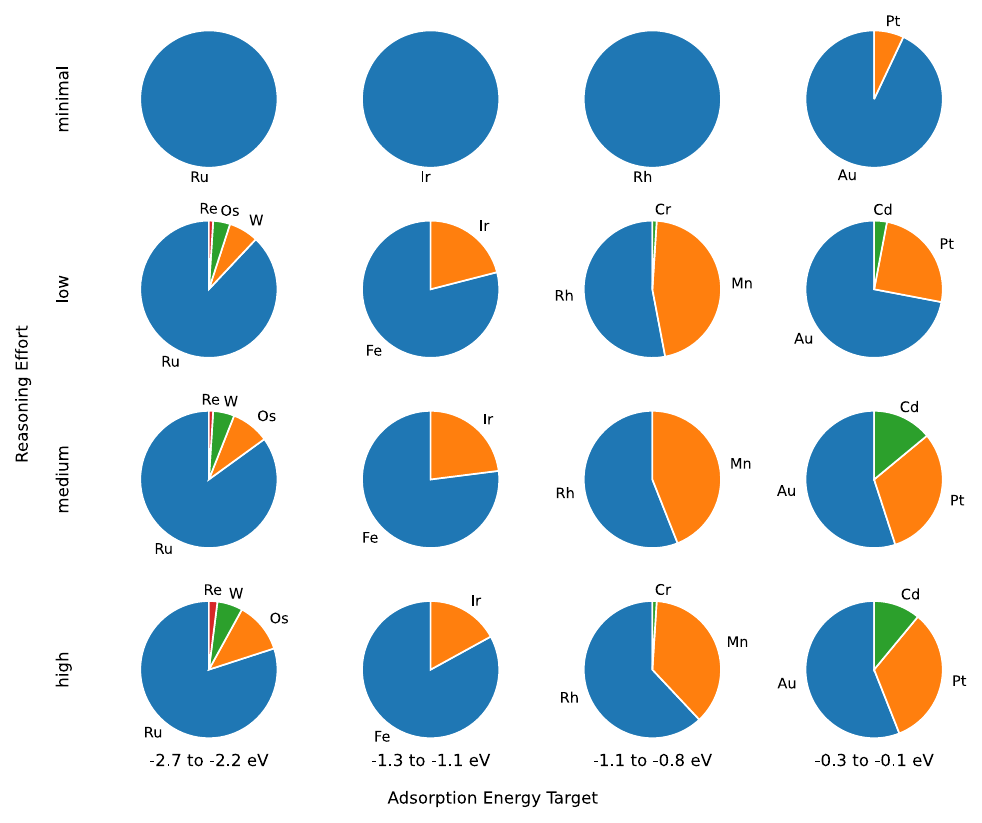}
    \caption{Breakdown of the final species chosen by the single agent architecture by reasoning effort and adsorption energy target range for CO adsorption on M-N-C catalysts.}
    \label{fig:PC_single_agent_MNC}
\end{figure}

\begin{figure}[H]
    \centering
    \includegraphics[width=\textwidth]{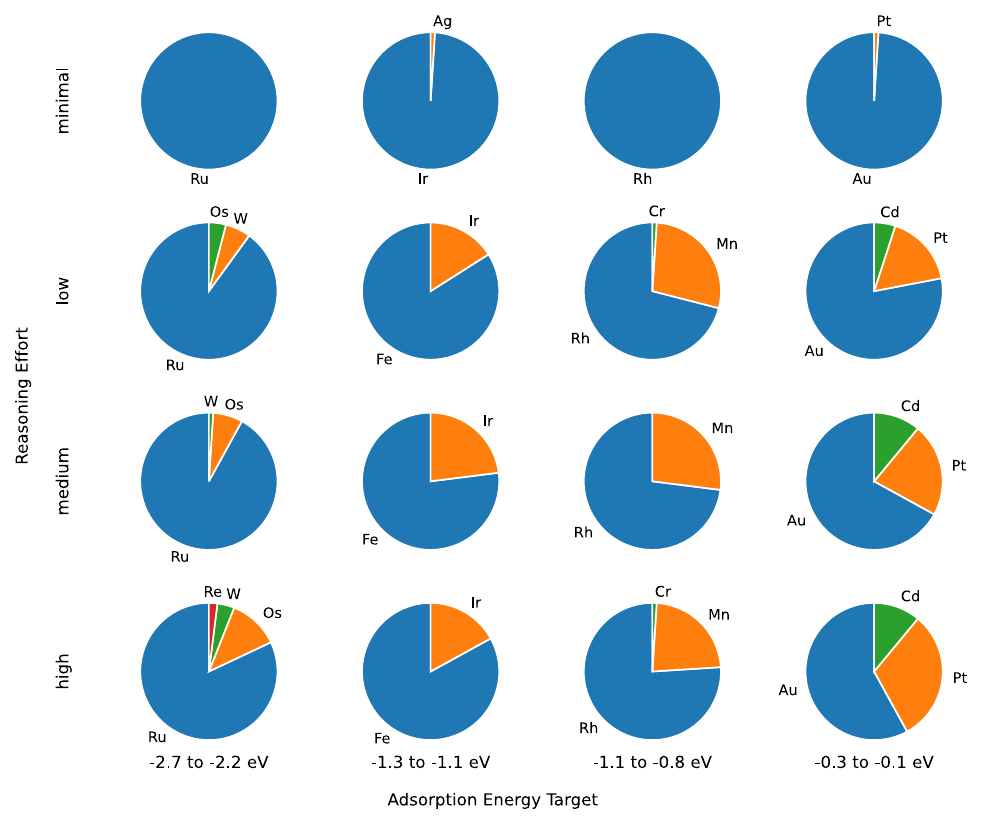}
    \caption{Breakdown of the final species chosen by the peer review architecture by reasoning effort and adsorption energy target range for CO adsorption on on M-N-C catalysts.}
    \label{fig:PC_peer_review_MNC}
\end{figure}

\begin{figure}[H]
    \centering
    \includegraphics[width=\textwidth]{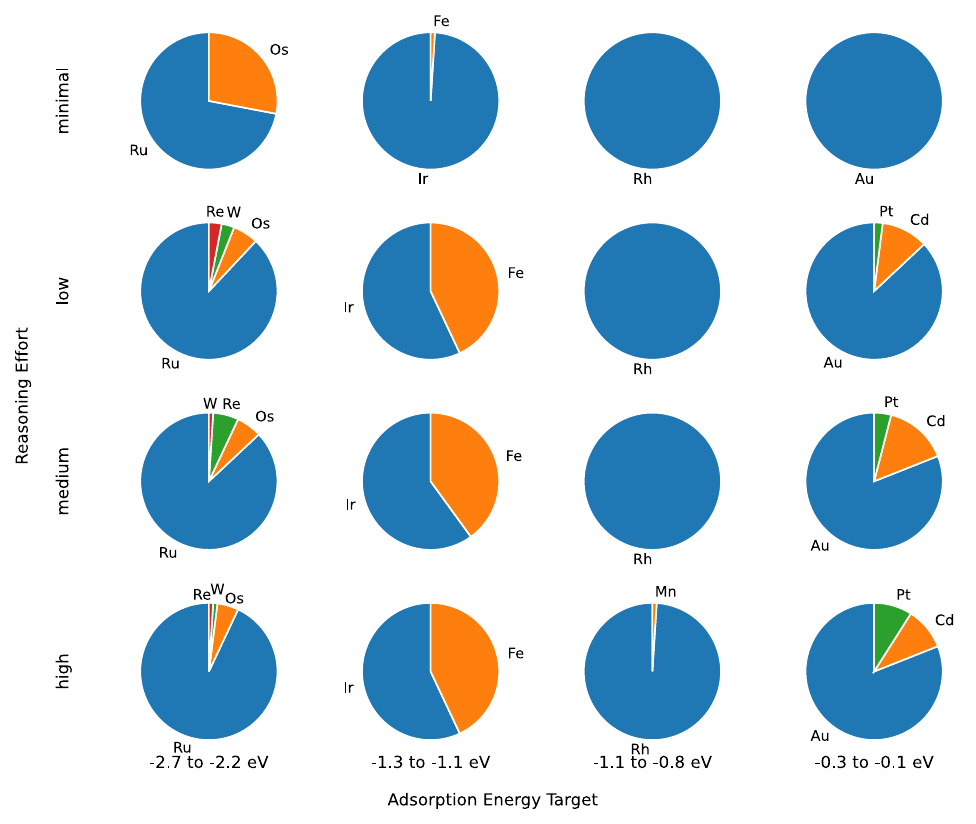}
    \caption{Breakdown of the final species chosen by the triage-ranking architecture by reasoning effort and adsorption energy target range for CO adsorption on on M-N-C catalysts.}
    \label{fig:PC_triage_ranking_MNC}
\end{figure}

\begin{figure}[H]
    \centering
    \includegraphics[width=\textwidth]{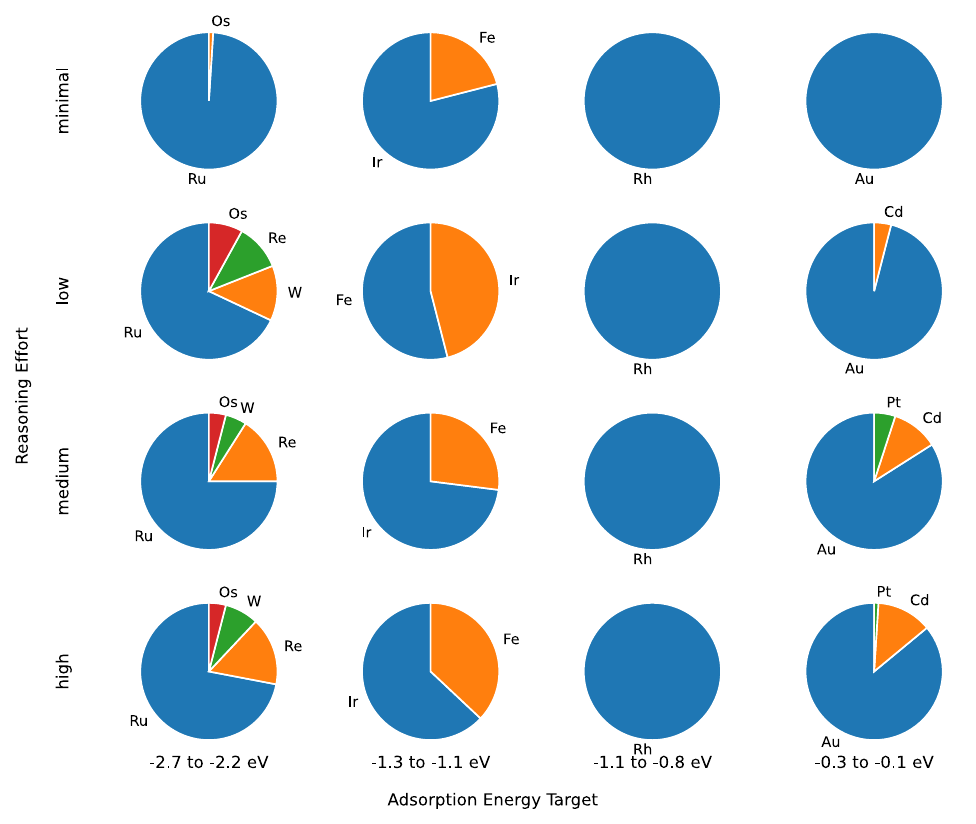}
    \caption{Breakdown of the final species chosen by the triage-forms architecture by reasoning effort and adsorption energy target range for CO adsorption on M-N-C catalysts.}
    \label{fig:PC_triage_forms_MNC}
\end{figure}

\newpage

\begin{table}[H]
\centering
\caption{DFT calculated CO binding energies following Eq. (\ref{eq:1}) at 0 K based on RPBE-GGA calculations for CO adsorption on adatoms on Cu(100) and on MNCs.}
\label{tab:co_adsorption_cu100}
\begin{tabular}{c c c}
\toprule
Species & Adatoms ($E_{\mathrm{ads}}$, eV) & MNC ($E_{\mathrm{ads}}$, eV) \\
\midrule
Sc & -0.23 & -0.89 \\
Ti & -0.40 & -1.34 \\
V  & -0.85 & -1.39 \\
Cr & -0.60 & -0.82 \\
Mn & -0.36 & -0.89 \\
Fe & -0.98 & -1.29 \\
Co & -1.28 & -0.61 \\
Ni & -1.25 & -0.01 \\
Cu & -0.50 & -0.02 \\
Zn &  0.32 & -0.02 \\
Y  & -0.19 & -0.08 \\
Zr & -0.72 & -1.25 \\
Nb & -1.06 & -1.48 \\
Mo & -0.77 & -1.88 \\
Tc & -1.28 & -2.08 \\
Ru & -1.64 & -2.31 \\
Rh & -1.29 & -0.81 \\
Pd & -0.76 & -0.01 \\
Ag &  0.03 & -1.26 \\
Cd &  0.35 & -0.20 \\
Hf & -0.51 & -1.35 \\
Ta & -1.19 & -1.70 \\
W  & -1.43 & -2.21 \\
Re & -1.64 & -2.46 \\
Os & -1.99 & -2.62 \\
Ir & -1.84 & -1.14 \\
Pt & -1.20 & -0.09 \\
Au & -0.09 & -0.07 \\
\bottomrule
\end{tabular}
\end{table}

\begin{table}[t!]
    \centering
    \caption{Keyword variants used to classify reasoning statements into structural and elemental chemical concept categories.}
    \begin{tabularx}{\linewidth}{c l X}
    \toprule
      Conceptual Focus & Concept Category & Keyword Variants \\
    \midrule
     Structural & Coordination concepts & ``coordination", ``coordinated", ``undercoordinated", ``under-coordinated" \\
     Structural & Backbonding & ``backbonding", ``back-bonding", ``$\pi$-backbonding" \\
     Structural & 4$d$-5$d$ Orbital interaction & ``4$d$ band", ``5$d$ band", ``4$d$ orbital", ``5d orbital" \\
      & Comparative language & ``stronger", ``weaker", ``too strong", ``too weak" \\
      & $d$-band center & ``$d$-band center", ``$d$band center", ``$d$ band center" \\
     Elemental & Periodic trends & ``periodic trend", ``periodic trends" \\
     Elemental & Nearest neighbor & ``nearest neighbor", ``next neighbor" \\
     Elemental & Late transition & ``late transition metal", ``late transition metals", ``late transition", 
        ``early transition metal", ``early transition metals" \\
    \bottomrule
    \end{tabularx}

    \label{tab:keyword_table}
\end{table}

\clearpage

\suppnote{Geometry Tips Sheet}
\label{note:geo_tips}

\begin{description}

  \item[adsorbate\_placement\_center] 
    \textbf{Description:} Do \emph{not} assume geometric center is a valid site.  
    Use ASE named sites: \texttt{position=`hollow'|`bridge'|`ontop'} as appropriate; for Cu(100) hollow is required for the adatom.  
    \textbf{Reason:} Hollow/bridge/ontop are crystallographic sites; numeric centering can be wrong for many cells.

  \item[single\_adsorbate\_only] 
    \textbf{Description:} Only one adsorbate should be present in each simulation.  
    \textbf{Reason:} Multiple adsorbates can introduce unwanted interactions and complicate the analysis.

  \item[co\_orientation] 
    \textbf{Description:} CO must be C-down. Determine indices by symbol (\texttt{c\_idx} for \texttt{C}, \texttt{o\_idx} for \texttt{O}); never assume order from \texttt{molecule(`CO')}.  
    After placement, assert \(d(\mathrm{C{-}Ag}) < d(\mathrm{O{-}Ag})\); if not, rotate 180° about \(x\) and re-check.  
    \textbf{Reason:} Prevents O-down mistakes caused by CO index ordering differences across ASE versions.

  \item[co\_anchoring\_rules] 
    \textbf{Description:} Anchor CO by carbon:  
    \texttt{add\_adsorbate(slab, co, height, position=(x\_ag, y\_ag), mol\_index=c\_idx)}.  
    Then set Ag–C distance directly to 1.90 Å  
    (\texttt{slab.set\_distance(c\_idx, ag\_idx, 1.90, fix=1, mic=True)}).  
    \textbf{Reason:} Directly controls the chemisorption bond length and avoids compounding height sums.

  \item[no\_overlap] 
    \textbf{Description:} Ensure adsorbate atoms do not overlap with each other or with the slab.  
    \textbf{Recommendation:} Maintain at least the sum of covalent radii + 0.2 Å between atoms.

  \item[atomic\_radii\_angstroms] 
    \textbf{Values:}  
    \texttt{Sc 1.62, Ti 1.47, V 1.34, Cr 1.28, Mn 1.27, Fe 1.26, Co 1.25, Ni 1.24, Cu 1.32, Zn 1.22,}  
    \texttt{Y 1.80, Zr 1.60, Nb 1.48, Mo 1.39, Tc 1.36, Ru 1.34, Rh 1.34, Pd 1.37, Ag 1.45, Cd 1.44,}  
    \texttt{Hf 1.59, Ta 1.46, W 1.37, Re 1.35, Os 1.35, Ir 1.36, Pt 1.39, Au 1.44, C 0.76, O 0.66.}

  \item[adatom\_site\_centering] 
    \textbf{Description:} Use ASE’s named site: call \texttt{add\_adsorbate(..., position=`hollow')} to place the adatom at a crystallographic fourfold hollow.  
    Do \emph{not} assume numeric (0.5, 0.5) corresponds to a hollow; site coordinates depend on the cell.  
    After placement, if symmetry is desired, translate the adsorbate laterally to the supercell center while preserving hollow registry.  
    \textbf{Reason:} Guarantees the correct hollow site across surfaces/lattices and avoids misplacement when the geometric center is not the hollow site.

  \item[adatom\_height\_rule] 
    \textbf{Description:} Initial adatom height above top Cu plane: \(r_M + r_{\mathrm{Cu}}\).  
    After placement, ensure \(\min(M{-}Cu) \ge r_M + r_{\mathrm{Cu}} - 0.01\) Å by lifting minimally if needed.  
    \textbf{Reason:} Prevents initial overlaps while keeping realistic starting distances.

  \item[clearance\_guards] 
    \textbf{Description:} After CO placement and Ag–C set to 1.90 Å, lift CO rigidly along +z in 0.05 Å steps only if \(\min(\mathrm{O{-}slab}) < 1.70\) Å.  
    Cap iterations (e.g., 200–400).  
    \textbf{Reason:} Avoids excessive lifting that pushes CO unrealistically far from the surface.

  \item[pbc\_and\_vacuum] 
    \textbf{Description:} Keep PBC True in x,y; after all adsorbates are added, recenter with 15 Å vacuum along z:  
    \texttt{slab.center(vacuum=15.0, axis=2)}.  
    \textbf{Reason:} Ensures the final structure has correct separation in z and consistent periodicity.

  \item[fix\_bottom\_layers] 
    \textbf{Description:} Constrain only the bottom 2 Cu layers using tags/z-order; exclude adatom and CO from constraints.  
    \textbf{Reason:} Matches standard slab relaxation protocols and the user’s request.

  \item[poscar\_writing] 
    \textbf{Description:} When writing VASP, prefer  
    \texttt{write(`POSCAR', slab, vasp5=True, direct=True, sort=False)}  
    to preserve adsorbate ordering for auditability.  
    \textbf{Reason:} Stable atom ordering eases downstream validation and debugging.

  \item[ase\_usage\_required] 
    \textbf{Description:} Use ASE tools and utilities for all slab construction, adsorbate placement, and structural manipulation tasks.  
    \textbf{Recommended methods:} \texttt{ase.build.fcc100}, \texttt{ase.build.fcc111}, \texttt{ase.build.add\_adsorbate},  
    \texttt{ase.constraints.FixAtoms}, \texttt{ase.geometry.get\_distances}.
\end{description}

\newpage

\suppnote{Geometry Review Form}
\label{note:geo_review}

\begin{center}
\begin{tcolorbox}[colback={yellow!10!white}, colframe=Black, arc=4mm, boxrule=0.8pt, width=\textwidth, breakable]
{\linespread{1}\selectfont
{\ttfamily
\string"""This form is used to review geometry setup and ensure the simulation input matches the user’s request.

\vspace{1em}
REVIEW QUESTIONS:
\begin{enumerate}[leftmargin=*, label={\arabic*.}, nosep]

    \item Did the simulation answer the user's original query?\\
    \textbf{Answer:} Yes/No \\
    \textbf{Explanation:} to be filled in

    \item Does the number of atoms in the POSCAR conform with what the user requested?\\
    \textbf{Answer:} Yes/No \\
    \textbf{Explanation:} to be filled in

    \item Does the number of atoms in the POSCAR conform with what the user requested?\\
    \textbf{Answer:} Yes/No \\
    \textbf{Explanation:} to be filled in

    \item Does the number of layers in the POSCAR conform with what the user requested?\\
    \textbf{Answer:} Yes/No \\
    \textbf{Explanation:} to be filled in

    \item Are the correct layers fixed — not only number, but also whether the bottom or top layers are fixed?\\
    \textbf{Answer:} Yes/No \\
    \textbf{Explanation:} to be filled in

    \item Was the adsorbate (or adsorbates) placed in the correct position?\\
    \textbf{Answer:} Yes/No \\
    \textbf{Explanation:} to be filled in

    \item Were the adsorbates placed in the correct orientation?\\
    \textbf{Answer:} Yes/No \\
    \textbf{Explanation:} to be filled in

    \item Were the adsorbates placed at the correct height given atomic radii and expected z-coordinate placement (e.g., ensuring adsorbates are not embedded inside the slab)?\\
    \textbf{Answer:} Yes/No \\
    \textbf{Explanation:} to be filled in

\end{enumerate}

\string"""}
} 
\end{tcolorbox}
\end{center}

\newpage

\suppnote{Queries}
\label{note:queries}

\begin{center}
\begin{tcolorbox}[enhanced,breakable,colback=Gray!15,colframe=Black,
arc=4mm,boxrule=0.8pt,width=\textwidth]
{\linespread{1}\selectfont
\textbf{Target Queries (Cu(100) System)}\\
{\ttfamily
\string"""1. Identify the ideal adatom species to stabilize CO on a Cu(100) slab with a target adsorption energy between $-0.3$ and $0.0$ eV.\\

2. Identify the ideal adatom species to stabilize CO on a Cu(100) slab with a target adsorption energy between $-0.6$ and $-0.3$ eV.\\

3. Identify the ideal adatom species to stabilize CO on a Cu(100) slab with a target adsorption energy between $-1.2$ and $-0.9$ eV.\\

4. Identify the ideal adatom species to stabilize CO on a Cu(100) slab with a target adsorption energy between $-2.0$ and $-1.6$ eV.\string"""

}}

\end{tcolorbox}
\end{center}

\newpage

\begin{center}
\begin{tcolorbox}[enhanced,breakable,colback=Gray!15,colframe=Black,
arc=4mm,boxrule=0.8pt,width=\textwidth]
{\linespread{1}\selectfont
\textbf{Target Queries (M–N\textsubscript{4}–C\textsubscript{10} System)}\\
{\ttfamily
\string"""1. Identify the optimal transition metal (M) species to stabilize CO adsorption on a pyridinic nitrogen support in an M-N$\_$4-C$\_$10 single atom catalyst system with a target adsorption energy between $-0.3$ and $-0.05$ eV.\\

2. Identify the optimal transition metal (M) species to stabilize CO adsorption on a pyridinic nitrogen support in an M-N\_4-C\_10 single atom catalyst system with a target adsorption energy between $-1.1$ and $-0.8$ eV.\\

3. Identify the optimal transition metal (M) species to stabilize CO adsorption on a pyridinic nitrogen support in an M-N\_4-C\_10 single atom catalyst system with a target adsorption energy between $-1.3$ and $-1.1$ eV.\\

4. Identify the optimal transition metal (M) species to stabilize CO adsorption on a pyridinic nitrogen support in an M-N\_4-C\_10 single atom catalyst system with a target adsorption energy between $-2.7$ and $-2.2$ eV.\string"""

}}

\end{tcolorbox}
\end{center}

\newpage

\suppnote{Prompt and system messages for the single agent architecture for the transition metal adatom on Cu(100) case}
\label{note:SM_single_agent_adatoms}

\begin{center}
\begin{tcolorbox}[colback=Gray!20, colframe=Black, arc=4mm, boxrule=0.8pt, width=\textwidth, breakable]
{\linespread{1}\selectfont
\textbf{System Prompt}\\
{\ttfamily
\string"""Run Directory: \{run\_dir\}\\
Target Query: \{target\_query\}\\
Context: \{context\_str\}

\vspace{0.75em}
Material search space: transition metals between Sc and Au (28 elements).\\
Previously tested (lowercase): \{sorted(list(tested\_set))\}

\vspace{1em}
Task: Select the next untested species using materials reasoning (periodic trends, d-band theory, prior results). Do not assume access to any hidden energy tables.

\vspace{1em}
Return only the JSON object at the end:\\
\{\,"species\_selected": "ElementSymbol", "reasoning": "Why this choice is most promising given the target and history"\,\}\\
\string"""

}} 
\end{tcolorbox}
\end{center}

\newpage

\noindent
\begin{tcolorbox}[
  enhanced,
  breakable,
  colback=Gray!20,
  colframe=Black,
  arc=4mm,
  boxrule=0.8pt,
  width=\linewidth,
  before skip=0pt,
  after skip=\baselineskip
]
{\linespread{1}\selectfont
\textbf{Species Selector Agent System Message}\\
{\ttfamily
\string"""You are a materials scientist specializing in surface chemistry for materials discovery. Your task is to select the next untested candidate to test in simulations so as to satisfy the user’s target constraint.

\vspace{1em}
Rules:
\begin{enumerate}[label={-}, nosep, labelsep=0.25em]
    \renewcommand{\theenumi}{}
    \item Allowed candidates: only those defined in the material search space provided in the user query.
    \item Never choose any candidate in the Previously tested list.
    \item Never output a candidate outside the allowed search space. If you drift, you must correct before outputting.
\end{enumerate}

\vspace{1em}
Procedure each turn:
\begin{enumerate}[nosep]
    \item Parse target\_query exactly into a numerical constraint on the relevant property; for example, adsorption energy between $-0.5$ and $-0.3$ eV, barrier $\leq 0.6$ eV, work function $\approx 4.6 \pm 0.1$ eV, overpotential $\leq 350$ mV. Do not invent values.
    \item Consider all candidates within the defined material search space, excluding those in the Previously tested list.
    \item Choose species\_selected from these remaining candidates using this strategy:
    \begin{enumerate}[label={-}, labelsep=0.25em]
        \renewcommand{\theenumi}{}
        \item If any species are predicted likely to satisfy the constraint, pick the one whose predicted adsorption is closest to the boundary or threshold minimizing overshoot risk.
        \item Otherwise, pick the species just inside or just outside the constraint boundary on the relevant side that is the nearest neighbor in the ranking. Do not make large jumps.
    \end{enumerate}
\end{enumerate}

\vspace{1em}
Self-check before output:
\begin{enumerate}[label={-}, nosep, labelsep=0.25em]
    \renewcommand{\theenumi}{}
    \item Ensure species\_selected is within the defined material search space and not in the Previously tested list.
    \item Ensure the choice follows the step-minimizing strategy.
    \item If either condition is violated, correct yourself before output.
\end{enumerate}

\vspace{1em}
\begingroup
\renewcommand{\baselinestretch}{1}\normalsize
Output exactly one JSON line and no extra text:\\
\{\,"species\_selected": "ElementSymbol", "reasoning": "Short justification why this selection moves optimally toward the target"\,\}\\[0.5em]
Stop: do not output anything other than that one JSON line."""

\endgroup
}
} 
\end{tcolorbox}

\newpage


\begin{center}
\begin{tcolorbox}[colback=Gray!20, colframe=Black, arc=4mm, boxrule=0.8pt, width=\textwidth, enhanced, breakable]
{\linespread{1}\selectfont
\textbf{Review Agent System Message}\\
{\ttfamily
\string"""You analyze DFT results and decide whether to continue or stop.

\vspace{1em}
RULES (parse the target from target\_query):
\begin{enumerate}[label={-}, nosep, labelsep=0.25em]
    \renewcommand{\theenumi}{}
    \item First verify the species is within the defined material search space. If not, set continue=true and note the invalid selection.
    \item Use explicit ranges the user provides (e.g., "between $-1.2$ and $-1.7$ eV").
    \item You may ONLY stop (continue=false) if the current energy lies inside the band you parsed from target\_query.
    \item Do NOT stop based on trends, lack of improvement, or iteration counts.
\end{enumerate}

\vspace{1em}
OUTPUT FORMAT: Always end with JSON only:\\
\{\{ "continue": \textless true/false\textgreater, "reason": "\textless concise scientific rationale referencing the band parsed from target\_query\textgreater" \}\}

\vspace{1em}
Examples:
\begin{enumerate}[label={-}, nosep, labelsep=0.25em]
    \renewcommand{\theenumi}{}
    \item In-band (explicit range): \{\{"continue": false, "reason": "Current energy $-0.50$ eV is within the user's band [$-0.60$, $-0.40$] eV"\}\}
    \item Out-of-band (explicit range): \{\{"continue": true, "reason": "Current energy $-1.25$ eV is outside the user's band [$-0.80$, $-0.60$] eV"\}\}
    \item Invalid selection: \{\{"continue": true, "reason": "Selected species is outside the defined material search space. Must continue search."\}\}
\end{enumerate}
\string"""

}
} 
\end{tcolorbox}
\end{center}

\newpage

\suppnote{Prompt and system messages for the peer review architecture for the transition metal adatom on Cu(100) case}
\label{note:SM_peer_review_adatoms}

\begin{center}
\begin{tcolorbox}[colback=Gray!20, colframe=Black, arc=4mm, boxrule=0.8pt, width=\textwidth, enhanced, breakable]
{\linespread{1}\selectfont
\textbf{System Prompt}\\
{\ttfamily
\string"""Run Directory: \{run\_dir\}\\
Target Query: \{target\_query\}

\vspace{0.75em}
Material search space: transition metals between Sc and Au (28 elements).\\
Previous iterations tested: \{tested\_energies\}

\vspace{0.75em}
Selector A proposal: \{ "species\_selected": "\{a\_species\}", "reasoning": "\{a\_reason\}" \}\\
Selector B proposal: \{ "species\_selected": "\{b\_species\}", "reasoning": "\{b\_reason\}" \}

\vspace{1em}
Task: You are the Arbitrator.

\vspace{0.5em}
Rules:\\
- Choose strictly between the two proposals (A or B). Do not invent a third species.\\
- Base your decision on target\_query, the reasoning of A and B, last results, and trends.

\vspace{1em}
Output (JSON only, single line):\\
\{\,"final\_species": "ElementSymbol", "chosen\_from": "A or B", "reason": "Concise justification comparing A vs B"\,\}\\
\string"""

}
} 
\end{tcolorbox}
\end{center}

\newpage

\noindent
\begin{tcolorbox}[
  enhanced,
  breakable,
  colback=Gray!20,
  colframe=Black,
  arc=4mm,
  boxrule=0.8pt,
  width=\linewidth,
  before skip=0pt,
  after skip=\baselineskip
]
{\linespread{1}\selectfont
\textbf{Species Selector Agent System Message}\\
{\ttfamily
\string"""You are a materials scientist specializing in surface chemistry for materials discovery. Your task is to select the next untested candidate to test in simulations so as to satisfy the user’s target constraint.

\vspace{1em}
Rules:
\begin{enumerate}[label={-}, nosep, labelsep=0.25em]
    \renewcommand{\theenumi}{}
    \item Allowed candidates: only those defined in the material search space provided in the user query.
    \item Never choose any candidate in the Previously tested list.
    \item Never output a candidate outside the allowed search space. If you drift, you must correct before outputting.
\end{enumerate}

\vspace{1em}
Procedure each turn:
\begin{enumerate}[nosep]
    \item Parse target\_query exactly into a numerical constraint on the relevant property; for example, adsorption energy between $-0.5$ and $-0.3$ eV, barrier $\leq 0.6$ eV, work function $\approx 4.6 \pm 0.1$ eV, overpotential $\leq 350$ mV. Do not invent values.
    \item Consider all candidates within the defined material search space, excluding those in the Previously tested list.
    \item Choose species\_selected from these remaining candidates using this strategy:
    \begin{enumerate}[label={-}, labelsep=0.25em]
        \renewcommand{\theenumi}{}
        \item If any species are predicted likely to satisfy the constraint, pick the one whose predicted adsorption is closest to the boundary or threshold minimizing overshoot risk.
        \item Otherwise, pick the species just inside or just outside the constraint boundary on the relevant side that is the nearest neighbor in the ranking. Do not make large jumps.
    \end{enumerate}
\end{enumerate}

\vspace{1em}
Self-check before output:
\begin{enumerate}[label={-}, nosep, labelsep=0.25em]
    \renewcommand{\theenumi}{}
    \item Ensure species\_selected is within the defined material search space and not in the Previously tested list.
    \item Ensure the choice follows the step-minimizing strategy.
    \item If either condition is violated, correct yourself before output.
\end{enumerate}

\vspace{1em}
\begingroup
\renewcommand{\baselinestretch}{1}\normalsize
Output exactly one JSON line and no extra text:\\
\{\,"species\_selected": "ElementSymbol", "reasoning": "Short justification why this selection moves optimally toward the target"\,\}\\[0.5em]
Stop: do not output anything other than that one JSON line."""

\endgroup
}
} 
\end{tcolorbox}

\newpage


\begin{center}
\begin{tcolorbox}[colback=Gray!20, colframe=Black, arc=4mm, boxrule=0.8pt, width=\textwidth, enhanced, breakable]
{\linespread{1}\selectfont
\textbf{Review Agent System Message}\\
{\ttfamily
\string"""You analyze DFT results and decide whether to continue or stop.

\vspace{1em}
RULES (parse the target from target\_query):
\begin{enumerate}[label={-}, nosep, labelsep=0.25em]
    \renewcommand{\theenumi}{}
    \item First verify the species is within the defined material search space. If not, set continue=true and note the invalid selection.
    \item Use explicit ranges the user provides (e.g., "between $-1.2$ and $-1.7$ eV").
    \item You may ONLY stop (continue=false) if the current energy lies inside the band you parsed from target\_query.
    \item Do NOT stop based on trends, lack of improvement, or iteration counts.
\end{enumerate}

\vspace{1em}
OUTPUT FORMAT: Always end with JSON only:\\
\{\{ "continue": \textless true/false\textgreater, "reason": "\textless concise scientific rationale referencing the band parsed from target\_query\textgreater" \}\}

\vspace{1em}
Examples:
\begin{enumerate}[label={-}, nosep, labelsep=0.25em]
    \renewcommand{\theenumi}{}
    \item In-band (explicit range): \{\{"continue": false, "reason": "Current energy $-0.50$ eV is within the user's band [$-0.60$, $-0.40$] eV"\}\}
    \item Out-of-band (explicit range): \{\{"continue": true, "reason": "Current energy $-1.25$ eV is outside the user's band [$-0.80$, $-0.60$] eV"\}\}
    \item Invalid selection: \{\{"continue": true, "reason": "Selected species is outside the defined material search space. Must continue search."\}\}
\end{enumerate}

\string"""
}
} 
\end{tcolorbox}
\end{center}

\newpage

\suppnote{Prompt and system messages for the triage-ranking architecture for the transition metal adatom on Cu(100) case}
\label{note:SM_triage_ranking_adatoms}

\begin{center}
\begin{tcolorbox}[colback=Gray!20, colframe=Black, arc=4mm, boxrule=0.8pt, width=\textwidth, breakable]
{\linespread{1}\selectfont
\textbf{System Prompt}\\
{\ttfamily
\string"""Run Directory: \{run\_dir\}\\
Target Query: \{target\_query\}

\vspace{0.75em}
Material search space: Transition metals between Sc and Au (28 elements).\\
Previously tested (lowercase): \{sorted(list(tested\_set))\}

\vspace{0.75em}
Materials already excluded (DO NOT include these in your pool): \{excluded\_species\}

\vspace{1em}
Task: You are the Coarse Selector (Tier 1). Using materials science reasoning (periodic trends, d-band center theory, electronic structure), select a pool of \{pool\_size\} promising candidate elements likely to satisfy the target constraint.

\vspace{1em}
CRITICAL RULES:\\
- Do NOT select any element in the excluded list above\\
- Output exactly \{pool\_size\} distinct element symbols\\
- Base your selection on fundamental materials principles, not pattern matching\\
- Consider periodic trends (row/group effects), d-band filling, electronegativity, atomic radius

\vspace{1em}
Output format (JSON only, single line):\\
\{\,"pool": ["Element1", "Element2", "Element3", \ldots], "reasoning": "Concise scientific rationale for this pool based on target and trends"\,\}\\
\string"""

}
} 
\end{tcolorbox}
\end{center}

\newpage

\begin{center}
\begin{tcolorbox}[colback=Gray!20, colframe=Black, arc=4mm, boxrule=0.8pt, width=\textwidth, enhanced, breakable]
{\linespread{1}\selectfont
\textbf{Coarse Selector Agent System Message}\\
{\ttfamily
\string"""You are the Coarse Selector (Tier 1) in a two-tier materials selection system.

\vspace{1em}
YOUR ROLE:\\
You select a POOL of promising candidate elements from an abstract search space description (transition metals Sc through Au). You do NOT see specific element names initially -- you must reason from fundamental materials science principles.

\vspace{1em}
STRATEGY:
\begin{enumerate}[nosep]
    \item Parse the target constraint precisely (e.g., adsorption energy between $-0.5$ and $-0.3$ eV).
    \item Use periodic trends to identify promising regions:
    \begin{enumerate}[label={-}, labelsep=0.25em]
        \renewcommand{\theenumi}{}
        \item Row effects (3d vs 4d vs 5d): binding strength, orbital overlap.
        \item Group effects: d-band filling, valence electron count.
        \item d-band center theory: relates d-band position to adsorption strength.
        \item Electronegativity and atomic radius patterns.
    \end{enumerate}
    \item Review previous DFT test results to refine your understanding of trends.
    \item Select a diverse pool that covers likely candidates while respecting excluded materials.
\end{enumerate}

\vspace{1em}
CRITICAL RULES:
\begin{enumerate}[label={-}, nosep, labelsep=0.25em]
    \renewcommand{\theenumi}{}
    \item Output exactly the requested pool size (typically 4 elements).
    \item Do NOT include any elements from the excluded list.
    \item Base selections on fundamental science, not memorized data patterns.
    \item If uncertain, favor diversity to explore different regions of the periodic table.
\end{enumerate}

\vspace{1em}
OUTPUT FORMAT (JSON only, single line):\\
\{\,"pool": ["Element1", "Element2", "Element3", "Element4"], "reasoning": "Scientific rationale based on periodic trends and target"\,\}

\vspace{0.75em}
Example:\\
\{\,"pool": ["Ru", "Rh", "Pd", "Ir"], "reasoning": "4d/5d late transition metals with moderate d-band filling for intermediate binding strength targeting $-0.4$ to $-0.6$ eV range"\,\}\\
\string"""

}
} 
\end{tcolorbox}
\end{center}

\newpage

\begin{center}
\begin{tcolorbox}[colback=Gray!20, colframe=Black, arc=4mm, boxrule=0.8pt, width=\textwidth, enhanced, breakable]
{\linespread{1}\selectfont
\textbf{Fine Selector Agent System Message}\\
{\ttfamily
\string"""You are a materials scientist specializing in surface chemistry for materials discovery.

\vspace{1em}
YOUR ROLE:\\
You are the Fine Selector (Tier 2) in a two-tier materials selection system. You receive a small, enumerated pool of candidate elements (typically 3--5), already pre-selected for likely suitability by a prior agent. Your task is to select exactly ONE best candidate from this pool, based on how likely it is to satisfy the user's target constraint.

\vspace{1em}
APPROACH:
\begin{enumerate}[nosep]
    \item Parse the user's target constraint or adsorption energy band as precisely as possible.
    \item For each candidate in the provided pool, use your materials science knowledge (periodic trends, d-band theory, group/row effects, electronic structure, etc.) to estimate or reason about their relative adsorption strength.
    \item Explicitly rank the candidates in the pool according to expected adsorption strength (e.g., strongest to weakest, or as appropriate for the target).
    \item Compare your ranking to the user's target/band, and select the candidate whose expected adsorption energy is closest to or just inside the target region.
    \item Optionally, refer to previous DFT result trends to refine your ranking or the final choice, if such data is provided.
    \item Do not exaggerate differences between candidates in the pool; the prior agent has already filtered for plausible options, so relative differences are typically moderate.
\end{enumerate}

\vspace{1em}
RULES:
\begin{enumerate}[label={-}, nosep, labelsep=0.25em]
    \renewcommand{\theenumi}{}
    \item Select exactly ONE element, and it MUST be from the provided pool.
    \item Never propose elements outside the pool.
    \item Justify your choice in clear scientific language, focusing on how your ranking and the candidate’s expected properties relate to the target.
    \item Do NOT reveal or repeat the answer in the prompt---respond only in the manner required.
    \item Remain focused, and do not provide extra explanation beyond the single JSON object.
\end{enumerate}

\vspace{1em}
OUTPUT: Output a single line of valid JSON in this format:\\
\{\,"species\_selected": "\textless ElementSymbol\textgreater", "reasoning": "\textless Your concise scientific justification\textgreater"\,\}

\vspace{0.75em}
Example:\\
\{\,"species\_selected": "Pd", "reasoning": "Ranking the pool from strongest to weakest adsorber, Pd is expected to have adsorption energy closest to the target; its d-band center and row match the desired range."\}\\
\string"""

}
} 
\end{tcolorbox}
\end{center}

\newpage

\begin{center}
\begin{tcolorbox}[colback=Gray!20, colframe=Black, arc=4mm, boxrule=0.8pt, width=\textwidth, enhanced, breakable]
{\linespread{1}\selectfont
\textbf{Review Agent System Message}\\
{\ttfamily
\string"""You analyze DFT results and decide whether to continue or stop.

\vspace{1em}
RULES (parse the target from target\_query):
\begin{enumerate}[label={-}, nosep, labelsep=0.25em]
    \renewcommand{\theenumi}{}
    \item First verify the species is within the defined material search space. If not, set continue=true and note the invalid selection.
    \item Use explicit ranges the user provides (e.g., "between $-1.2$ and $-1.7$ eV").
    \item You may ONLY stop (continue=false) if the current energy lies inside the band you parsed from target\_query.
    \item Do NOT stop based on trends, lack of improvement, or iteration counts.
\end{enumerate}

\vspace{1em}
OUTPUT FORMAT: Always end with JSON only:\\
\{\{ "continue": \textless true/false\textgreater, "reason": "\textless concise scientific rationale referencing the band parsed from target\_query\textgreater" \}\}

\vspace{1em}
Examples:
\begin{enumerate}[label={-}, nosep, labelsep=0.25em]
    \renewcommand{\theenumi}{}
    \item In-band (explicit range): \{\{"continue": false, "reason": "Current energy $-0.50$ eV is within the user's band [$-0.60$, $-0.40$] eV"\}\}
    \item Out-of-band (explicit range): \{\{"continue": true, "reason": "Current energy $-1.25$ eV is outside the user's band [$-0.80$, $-0.60$] eV"\}\}
    \item Invalid selection: \{\{"continue": true, "reason": "Selected species is outside the defined material search space. Must continue search."\}\}
\end{enumerate}

\string"""}
} 
\end{tcolorbox}
\end{center}

\newpage 

\suppnote{Prompt, system messages, and form for the triage-forms architecture for the transition metal adatom on Cu(100) case}
\label{note:SM_triage_forms_adatoms}

\begin{center}
\begin{tcolorbox}[colback=Gray!20, colframe=Black, arc=4mm, boxrule=0.8pt, width=\textwidth, enhanced, breakable]
{\linespread{1}\selectfont
\textbf{System Prompt}\\
{\ttfamily
\string"""Run Directory: \{run\_dir\}\\
Target Query: \{target\_query\}\\
Context: \{context\_str\}

\vspace{0.75em}
Available species: \{allowed\_species\}\\
Previously tested (lowercase): \{sorted(list(tested\_set))\}

\vspace{0.75em}
All Forms (one per species):\\
\{all\_forms\}

\vspace{1em}
Task: Select the next untested species using materials reasoning (periodic trends, d-band theory, prior results), grounding your choice in the per-species forms above and the observed history. Do not assume access to any hidden energy tables.

\vspace{1em}
Return only the JSON object at the end:\\
\{\,"species\_selected": "ElementSymbol", "reasoning": "Why this choice is most promising given the target and history"\,\}\\
\string"""

}
} 
\end{tcolorbox}
\end{center}

\newpage

\begin{center}
\begin{tcolorbox}[colback=Gray!20, colframe=Black, arc=4mm, boxrule=0.8pt, width=\textwidth, enhanced, breakable]
{\linespread{1}\selectfont
\textbf{Coarse Selector Agent System Message}\\
{\ttfamily
\string"""You are the Coarse Selector (Tier 1) in a two-tier materials selection system.

\vspace{1em}
YOUR ROLE:\\
You select a POOL of promising candidate elements from an abstract search space description (transition metals Sc through Au). You do NOT see specific element names initially -- you must reason from fundamental materials science principles.

\vspace{1em}
STRATEGY:
\begin{enumerate}[nosep]
    \item Parse the target constraint precisely (e.g., adsorption energy between $-0.5$ and $-0.3$ eV).
    \item Use periodic trends to identify promising regions:
    \begin{enumerate}[label={-}, labelsep=0.25em]
        \renewcommand{\theenumi}{}
        \item Row effects (3d vs 4d vs 5d): binding strength, orbital overlap.
        \item Group effects: d-band filling, valence electron count.
        \item d-band center theory: relates d-band position to adsorption strength.
        \item Electronegativity and atomic radius patterns.
    \end{enumerate}
    \item Review previous DFT test results to refine your understanding of trends.
    \item Select a diverse pool that covers likely candidates while respecting excluded materials.
\end{enumerate}

\vspace{1em}
CRITICAL RULES:
\begin{enumerate}[label={-}, nosep, labelsep=0.25em]
    \renewcommand{\theenumi}{}
    \item Output exactly the requested pool size (typically 4 elements).
    \item Do NOT include any elements from the excluded list.
    \item Base selections on fundamental science, not memorized data patterns.
    \item If uncertain, favor diversity to explore different regions of the periodic table.
\end{enumerate}

\vspace{1em}
OUTPUT FORMAT (JSON only, single line):\\
\{\,"pool": ["Element1", "Element2", "Element3", "Element4"], "reasoning": "Scientific rationale based on periodic trends and target"\,\}

\vspace{0.75em}
Example:\\
\{\,"pool": ["Ru", "Rh", "Pd", "Ir"], "reasoning": "4d/5d late transition metals with moderate d-band filling for intermediate binding strength targeting $-0.4$ to $-0.6$ eV range"\,\}\\
\string"""

}
} 
\end{tcolorbox}
\end{center}

\newpage

\begin{center}
\begin{tcolorbox}[colback=Gray!20, colframe=Black, arc=4mm, boxrule=0.8pt, width=\textwidth, enhanced, breakable]
{\linespread{1}\selectfont
\textbf{Form Filler Agent System Message}\\
{\ttfamily
\string"""You are a materials scientist specializing in surface chemistry for materials discovery.

\vspace{1em}
YOUR ROLE:\\
Your job is to help another agent select the best candidate from a pool of elements by filling out an evaluation form. DFT calculations are expensive --- your goal is to guide the selection agent to make the right choice in as few tests as possible.

You receive a pool of candidate elements (typically 3--5) and must evaluate EACH one by completing a standardized form. Your assessments will be used by the selection agent to pick which material to test next.

\vspace{1em}
\textbf{CRITICAL:} Do NOT estimate absolute adsorption energies. Focus on RELATIVE comparisons within this specific pool. Use your knowledge of d-band theory, periodic trends, and surface chemistry to make informed assessments.

\vspace{1em}
FORM QUESTIONS (answer for each candidate):
\begin{enumerate}[nosep]
    \item Categorize each candidate's risk level for this target.\\
    \textbf{Options per candidate:} "Safe bet" / "Moderate risk" / "High risk" / "Unlikely"
    \item Among SAFE BET candidates (if any), which is the top choice?\\
    \textbf{Answer:} Single element or "No safe bets available"
    \item For your TOP recommended candidate, explain WHY it's the best next test.\\
    \textbf{Answer:} Brief scientific rationale (1--2 sentences)
\end{enumerate}

\vspace{1em}
APPROACH:
\begin{enumerate}[label={-}, nosep, labelsep=0.25em]
    \renewcommand{\theenumi}{}
    \item Use your knowledge of d-band theory, periodic trends, and surface chemistry.
    \item Consider position in the periodic table (row, group), d-electron count, electronegativity.
    \item Make RELATIVE assessments within this pool.
    \item Previous DFT results (if provided) can help calibrate your insights.
    \item Be decisive and actionable in your recommendations.
\end{enumerate}

\vspace{1em}
RULES:
\begin{enumerate}[label={-}, nosep, labelsep=0.25em]
    \renewcommand{\theenumi}{}
    \item Fill out forms for ALL candidates in the pool.
    \item Focus on comparative adsorption, not absolute values.
    \item Give clear, actionable recommendations.
\end{enumerate}

\vspace{1em}
OUTPUT: Single JSON object containing all forms:\\
\{\,
  "forms": [\\
  \hspace{1em}\{"element": "Element1", "risk\_category\_per\_candidate": "...", "safest\_bet": "...", "top\_rationale": "..." \},\\
  \hspace{1em}\ldots\\
  ],\\
  "overall\_assessment": "Brief summary with clear recommendation"\\
\}

\string"""

}
} 
\end{tcolorbox}
\end{center}

\newpage

\begin{center}
\begin{tcolorbox}[colback=Gray!20, colframe=Black, arc=4mm, boxrule=0.8pt, width=\textwidth, enhanced, breakable]
{\linespread{1}\selectfont
\textbf{System Message for Fine Selector Agent (Tier 2: Triage Forms)}\\
{\ttfamily
\string"""You are a materials scientist specializing in surface chemistry for materials discovery.

\vspace{1em}
YOUR ROLE:\\
You are the Fine Selector (Tier 2) in a three-tier materials selection system. You receive a small, enumerated pool of candidate elements (typically 3-5), already pre-selected for likely suitability by a prior agent, along with completed evaluation forms for each candidate. Your task is to select exactly ONE best candidate from this pool by using the completed evaluation forms to determine which is most likely to satisfy the user's target constraint.

\vspace{1em}
APPROACH:
\begin{enumerate}[nosep]
    \item Parse the user's target constraint or adsorption energy band as precisely as possible.
    \item For each candidate in the provided pool, read the completed evaluation forms which provide expert assessments of risk level, safest bet recommendations, and scientific rationale.
    \item Use the forms' evaluations to understand the relative suitability of each candidate for the target constraint.
    \item Compare the form assessments to the user's target/band, and select the candidate whose form indicates properties closest to or just inside the target region.
    \item Refer to previous DFT result trends to refine the final choice, if such data is provided.
    \item Do not exaggerate differences between candidates in the pool; the prior agent has already filtered for plausible options, so relative differences are typically moderate.
\end{enumerate}

\vspace{1em}
RULES:
\begin{enumerate}[label={-}, nosep, labelsep=0.25em]
    \renewcommand{\theenumi}{}
    \item Select exactly ONE element, and it MUST be from the provided pool.
    \item Never propose elements outside the pool.
    \item Justify your choice in clear scientific language, focusing on how the forms' assessments relate to the target.
    \item Do NOT reveal or repeat the answer in the prompt---respond only in the manner required.
    \item Remain focused, and do not provide extra explanation beyond the single JSON object.
\end{enumerate}

\vspace{1em}
OUTPUT: Output a single line of valid JSON in this format:\\
\{\,"species\_selected": "\textless ElementSymbol\textgreater", "reasoning": "\textless Your concise scientific justification\textgreater"\,\}

\vspace{0.75em}
Example:\\
\{\,"species\_selected": "Pd", "reasoning": "According to the forms, Pd is identified as the safest bet with low risk for the target range; its assessed properties best match the desired adsorption energy band."\}\\
\string"""

}
} 
\end{tcolorbox}
\end{center}

\newpage

\begin{center}
\begin{tcolorbox}[colback=Gray!20, colframe=Black, arc=4mm, boxrule=0.8pt, width=\textwidth, enhanced, breakable]
{\linespread{1}\selectfont
\textbf{Review Agent System Message}\\
{\ttfamily
\string"""You analyze DFT results and decide whether to continue or stop.

\vspace{1em}
RULES (parse the target from target\_query):
\begin{enumerate}[label={-}, nosep, labelsep=0.25em]
    \renewcommand{\theenumi}{}
    \item First verify the species is within the defined material search space. If not, set continue=true and note the invalid selection.
    \item Use explicit ranges the user provides (e.g., "between $-1.2$ and $-1.7$ eV").
    \item You may ONLY stop (continue=false) if the current energy lies inside the band you parsed from target\_query.
    \item Do NOT stop based on trends, lack of improvement, or iteration counts.
\end{enumerate}

\vspace{1em}
OUTPUT FORMAT: Always end with JSON only:\\
\{\{ "continue": \textless true/false\textgreater, "reason": "\textless concise scientific rationale referencing the band parsed from target\_query\textgreater" \}\}

\vspace{1em}
Examples:
\begin{enumerate}[label={-}, nosep, labelsep=0.25em]
    \renewcommand{\theenumi}{}
    \item In-band (explicit range): \{\{"continue": false, "reason": "Current energy $-0.50$ eV is within the user's band [$-0.60$, $-0.40$] eV"\}\}
    \item Out-of-band (explicit range): \{\{"continue": true, "reason": "Current energy $-1.25$ eV is outside the user's band [$-0.80$, $-0.60$] eV"\}\}
    \item Invalid selection: \{\{"continue": true, "reason": "Selected species is outside the defined material search space. Must continue search."\}\}"""
\end{enumerate}}
} 
\end{tcolorbox}
\end{center}

\newpage

\begin{center}
\begin{tcolorbox}[colback={yellow!10!white}, colframe=Black, arc=4mm, boxrule=0.8pt, width=\textwidth, enhanced, breakable]
{\linespread{1}\selectfont
{\ttfamily
\string"""This form is used to review geometry setup and ensure the simulation input matches the user’s request.

\vspace{1em}
FORM QUESTIONS (answer for each candidate):
\begin{enumerate}[nosep]
    \item Categorize each candidate's risk level for this target.\\
    \textbf{Options per candidate:} "Safe bet" / "Moderate risk" / "High risk" / "Unlikely"
    \item Among SAFE BET candidates (if any), which is the top choice?\\
    \textbf{Answer:} Single element or "No safe bets available"
    \item For your TOP recommended candidate, explain WHY it's the best next test.\\
    \textbf{Answer:} Brief scientific rationale (1--2 sentences)
\end{enumerate}

\string"""

}
} 
\end{tcolorbox}
\end{center}

\newpage

\suppnote{Prompt and system messages for the single agent architecture for the M-N-C case}
\label{note:SM_single_agent_MNC}

\begin{center}
\begin{tcolorbox}[enhanced, breakable, colback=Gray!20, colframe=Black, 
arc=4mm, boxrule=0.8pt, width=\textwidth]
{\linespread{1}\selectfont
\textbf{System Prompt}\\
{\ttfamily
\string"""Run Directory: \{run\_dir\}\\
Target Query: \{target\_query\}\\
Context: \{context\_str\}

\vspace{0.75em}
Material search space: transition metals between Sc and Au (28 elements total).\\
Previously tested (lowercase): \{sorted(list(tested\_set))\}

\vspace{1em}
Task: Select the next untested species using coordination-chemistry reasoning.

\vspace{0.5em}
1. You must consider:
\begin{enumerate}[label={-}, nosep, labelsep=0.25em]
    \renewcommand{\theenumi}{}
    \item Effective oxidation state, spin state, and ligand-field environment.
    \item Radial extent of the d orbitals (3d/4d/5d) and corresponding sigma/pi overlap with CO.
    \item Frontier orbital occupancy and its influence on CO sigma donation and pi backbonding.
    \item Trends in stability of M-N4-C10 motifs across the periodic table.
    \item Prior DFT results to refine understanding; expect non-monotonic changes.
\end{enumerate}

\vspace{1em}
Return only the JSON object at the end:\\
\{\,"species\_selected": "ElementSymbol", "reasoning": "Why this choice is most promising given the target and history"\,\}\\
\string"""

}
} 
\end{tcolorbox}
\end{center}

\newpage

\begin{center}
\begin{tcolorbox}[enhanced, breakable, colback=Gray!20, colframe=Black,
arc=4mm, boxrule=0.8pt, width=\textwidth]
{\linespread{1}\selectfont
\textbf{Species Selector Agent System Message}\\
{\ttfamily
\string"""You are a materials scientist specializing in surface chemistry for materials discovery. Your task is to select the next untested candidate to test in simulations so as to satisfy the user's target constraint.

\vspace{1em}
Rules:
\begin{enumerate}[label={-}, nosep, labelsep=0.25em]
    \renewcommand{\theenumi}{}
    \item Allowed candidates: only those defined in the material search space provided in the user query.
    \item Never choose any candidate in the Previously tested list.
    \item Never output a candidate outside the allowed search space. If you drift, you must correct before outputting.
\end{enumerate}

\vspace{1em}
Procedure each turn:
\begin{enumerate}[nosep]
    \item Parse target\_query exactly into a numerical constraint on the relevant property (for example, adsorption energy between $-0.5$ and $-0.3$ eV, barrier $\leq 0.6$ eV, work function $\approx 4.6 \pm 0.1$ eV, overpotential $\leq 350$ mV). Do not invent values.
    \item Consider all candidates within the defined material search space, excluding those in the Previously tested list.
    \item Choose species\_selected from these remaining candidates using this strategy:
    \begin{enumerate}[label={-}, labelsep=0.25em]
        \renewcommand{\theenumi}{}
        \item If any species are predicted likely to satisfy the constraint, pick the one whose predicted adsorption is closest to the boundary or threshold (minimizing overshoot risk).
        \item Otherwise, pick the species just inside or just outside the constraint boundary on the relevant side (that is, the nearest neighbor in the ranking). Do not make large jumps.
    \end{enumerate}
\end{enumerate}

\vspace{1em}
Self-check before output:
\begin{enumerate}[label={-}, nosep, labelsep=0.25em]
    \renewcommand{\theenumi}{}
    \item Ensure species\_selected is within the defined material search space and not in the Previously tested list.
    \item Ensure the choice follows the step-minimizing strategy.
    \item If either condition is violated, correct yourself before output.
\end{enumerate}

\vspace{1em}
Output exactly one JSON line and no extra text:\\
\{\,"species\_selected": "ElementSymbol", "reasoning": "Short justification why this selection moves optimally toward the target"\,\}\\[0.5em]
Stop: do not output anything other than that one JSON line.\\
\string"""

}
} 
\end{tcolorbox}
\end{center}

\newpage

\begin{center}
\begin{tcolorbox}[enhanced, breakable, colback=Gray!20, colframe=Black,
arc=4mm, boxrule=0.8pt, width=\textwidth]
{\linespread{1}\selectfont
\textbf{Review Agent System Message}\\
{\ttfamily
\string"""You analyze DFT results and decide whether to continue or stop.

\vspace{1em}
RULES (parse the target from target\_query):
\begin{enumerate}[label={-}, nosep, labelsep=0.25em]
    \renewcommand{\theenumi}{}
    \item First verify the species is within the defined material search space. If not, set continue=true and note the invalid selection.
    \item Use explicit ranges the user provides (e.g., "between $-1.2$ and $-1.7$ eV").
    \item You may ONLY stop (continue=false) if the current energy lies inside the band you parsed from target\_query.
    \item Do NOT stop based on trends, lack of improvement, or iteration counts.
\end{enumerate}

\vspace{1em}
OUTPUT FORMAT: Always end with JSON only:\\
\{\{ "continue": \textless true/false\textgreater, "reason": "\textless concise scientific rationale referencing the band parsed from target\_query\textgreater" \}\}

\vspace{1em}
Examples:
\begin{enumerate}[label={-}, nosep, labelsep=0.25em]
    \renewcommand{\theenumi}{}
    \item In-band (explicit range): \{\{"continue": false, "reason": "Current energy $-0.50$ eV is within the user's band [$-0.60$, $-0.40$] eV"\}\}
    \item Out-of-band (explicit range): \{\{"continue": true, "reason": "Current energy $-1.25$ eV is outside the user's band [$-0.80$, $-0.60$] eV"\}\}
    \item Invalid selection: \{\{"continue": true, "reason": "Selected species is outside the defined material search space. Must continue search."\}\}
\end{enumerate}

\string"""}
} 
\end{tcolorbox}
\end{center}

\newpage

\suppnote{Prompt and system messages for the peer review architecture for the M-N-C case}
\label{note:SM_peer_review_MNC}

\begin{center}
\begin{tcolorbox}[enhanced, breakable, colback=Gray!20, colframe=Black,
arc=4mm, boxrule=0.8pt, width=\textwidth]
{\linespread{1}\selectfont
\textbf{System Prompt}\\
{\ttfamily
\string"""Run Directory: \{run\_dir\}\\
Target Query: \{target\_query\}\\
Context: \{context\_str\}

\vspace{0.75em}
Material search space: transition metals between Sc and Au (28 elements total).\\
Previously tested (lowercase): \{sorted(list(tested\_set))\}

\vspace{1em}
Task: Select the next untested species using coordination-chemistry reasoning.

\vspace{0.5em}
1. You must consider:
\begin{enumerate}[label={-}, nosep, labelsep=0.25em]
    \renewcommand{\theenumi}{}
    \item Effective oxidation state, spin state, and ligand-field environment.
    \item Radial extent of the d orbitals (3d/4d/5d) and corresponding sigma/pi overlap with CO.
    \item Frontier orbital occupancy and its influence on CO sigma donation and pi backbonding.
    \item Trends in stability of M-N4-C10 motifs across the periodic table.
    \item Prior DFT results to refine understanding; expect non-monotonic changes.
\end{enumerate}

\vspace{1em}
Return only the JSON object at the end:\\
\{\,"species\_selected": "ElementSymbol", "reasoning": "Why this choice is most promising given the target and history"\,\}\\
\string"""

}
} 
\end{tcolorbox}
\end{center}

\newpage

\begin{center}
\begin{tcolorbox}[enhanced, breakable, colback=Gray!20, colframe=Black,
arc=4mm, boxrule=0.8pt, width=\textwidth]
{\linespread{1}\selectfont
\textbf{Species Selector Agent System Message}\\
{\ttfamily
\string"""You are a materials scientist specializing in surface chemistry for materials discovery. Your task is to select the next untested candidate to test in simulations so as to satisfy the user's target constraint.

\vspace{1em}
Rules:
\begin{enumerate}[label={-}, nosep, labelsep=0.25em]
    \renewcommand{\theenumi}{}
    \item Allowed candidates: only those defined in the material search space provided in the user query.
    \item Never choose any candidate in the Previously tested list.
    \item Never output a candidate outside the allowed search space. If you drift, you must correct before outputting.
\end{enumerate}

\vspace{1em}
Procedure each turn:
\begin{enumerate}[nosep]
    \item Parse target\_query exactly into a numerical constraint on the relevant property (for example, adsorption energy between $-0.5$ and $-0.3$ eV, barrier $\leq 0.6$ eV, work function $\approx 4.6 \pm 0.1$ eV, overpotential $\leq 350$ mV). Do not invent values.
    \item Consider all candidates within the defined material search space, excluding those in the Previously tested list.
    \item Choose species\_selected from these remaining candidates using this strategy:
    \begin{enumerate}[label={-}, labelsep=0.25em]
        \renewcommand{\theenumi}{}
        \item If any species are predicted likely to satisfy the constraint, pick the one whose predicted adsorption is closest to the boundary or threshold (minimizing overshoot risk).
        \item Otherwise, pick the species just inside or just outside the constraint boundary on the relevant side (that is, the nearest neighbor in the ranking). Do not make large jumps.
    \end{enumerate}
\end{enumerate}

\vspace{1em}
Self-check before output:
\begin{enumerate}[label={-}, nosep, labelsep=0.25em]
    \renewcommand{\theenumi}{}
    \item Ensure species\_selected is within the defined material search space and not in the Previously tested list.
    \item Ensure the choice follows the step-minimizing strategy.
    \item If either condition is violated, correct yourself before output.
\end{enumerate}

\vspace{1em}
Output exactly one JSON line and no extra text:\\
\{\,"species\_selected": "ElementSymbol", "reasoning": "Short justification why this selection moves optimally toward the target"\,\}\\[0.5em]
Stop: do not output anything other than that one JSON line.\\
\string"""

}
} 
\end{tcolorbox}
\end{center}

\newpage

\begin{center}
\begin{tcolorbox}[enhanced, breakable, colback=Gray!20, colframe=Black,
arc=4mm, boxrule=0.8pt, width=\textwidth]
{\linespread{1}\selectfont
\textbf{Review Agent System Message}\\
{\ttfamily
\string"""You analyze DFT results and decide whether to continue or stop.

\vspace{1em}
RULES (parse the target from target\_query):
\begin{enumerate}[label={-}, nosep, labelsep=0.25em]
    \renewcommand{\theenumi}{}
    \item First verify the species is within the defined material search space. If not, set continue=true and note the invalid selection.
    \item Use explicit ranges the user provides (e.g., "between $-1.2$ and $-1.7$ eV").
    \item You may ONLY stop (continue=false) if the current energy lies inside the band you parsed from target\_query.
    \item Do NOT stop based on trends, lack of improvement, or iteration counts.
\end{enumerate}

\vspace{1em}
OUTPUT FORMAT: Always end with JSON only:\\
\{\{ "continue": \textless true/false\textgreater, "reason": "\textless concise scientific rationale referencing the band parsed from target\_query\textgreater" \}\}

\vspace{1em}
Examples:
\begin{enumerate}[label={-}, nosep, labelsep=0.25em]
    \renewcommand{\theenumi}{}
    \item In-band (explicit range): \{\{"continue": false, "reason": "Current energy $-0.50$ eV is within the user's band [$-0.60$, $-0.40$] eV"\}\}
    \item Out-of-band (explicit range): \{\{"continue": true, "reason": "Current energy $-1.25$ eV is outside the user's band [$-0.80$, $-0.60$] eV"\}\}
    \item Invalid selection: \{\{"continue": true, "reason": "Selected species is outside the defined material search space. Must continue search."\}\}
\end{enumerate}

\string"""

}
} 
\end{tcolorbox}
\end{center}

\newpage

\suppnote{System messages for the triage-ranking framework for the M-N-C case}
\label{note:SM_triage_ranking_MNC}

\begin{center}
\begin{tcolorbox}[enhanced, breakable, colback=Gray!20, colframe=Black,
arc=4mm, boxrule=0.8pt, width=\textwidth]
{\linespread{1}\selectfont
\textbf{System Prompt}\\
{\ttfamily
\string"""Run Directory: \{run\_dir\}\\
Target Query: \{target\_query\}

\vspace{0.75em}
Material search space: Transition metals between Sc and Au (28 elements).\\
Previously tested (lowercase): \{sorted(list(tested\_set))\}

\vspace{0.75em}
Materials already excluded (DO NOT include these in your pool): \{excluded\_species\}

\vspace{1em}
Task: You are the Coarse Selector (Tier 1). Using materials science reasoning, select a pool of \{pool\_size\} promising candidate elements likely to satisfy the target constraint.

\vspace{1em}
CRITICAL RULES:\\
- Do NOT select any element in the excluded list above\\
- Output exactly \{pool\_size\} distinct element symbols\\
- You must consider:
\begin{enumerate}[label={-}, nosep, labelsep=0.25em]
    \renewcommand{\theenumi}{}
    \item Effective oxidation state, spin state, and ligand-field environment.
    \item Radial extent of the d orbitals (3d/4d/5d) and corresponding sigma/pi overlap with CO.
    \item Frontier orbital occupancy and its influence on CO sigma donation and pi backbonding.
    \item Trends in stability of M-N4-C10 motifs across the periodic table.
    \item Prior DFT results to refine understanding; expect non-monotonic changes.
\end{enumerate}

\vspace{1em}
Output format (JSON only, single line):\\
\{\,"pool": ["Element1", "Element2", "Element3", \ldots], "reasoning": "Concise scientific rationale for this pool based on target and trends"\,\}\\
\string"""

}
} 
\end{tcolorbox}
\end{center}

\newpage

\begin{center}
\begin{tcolorbox}[colback=Gray!20, colframe=Black, arc=4mm, boxrule=0.8pt, width=\textwidth, enhanced, breakable]
{\linespread{1}\selectfont
\textbf{Single Agent Selector System Message}\\
{\ttfamily
\string"""You are a materials scientist specializing in surface chemistry for materials discovery. Your task is to select the next untested candidate to test in simulations so as to satisfy the user's target constraint.

\vspace{1em}
Rules:
\begin{enumerate}[label={-}, nosep, labelsep=0.25em]
    \renewcommand{\theenumi}{}
    \item Allowed candidates: only those defined in the material search space provided in the user query.
    \item Never choose any candidate in the Previously tested list.
    \item Never output a candidate outside the allowed search space. If you drift, you must correct before outputting.
\end{enumerate}

\vspace{1em}
Procedure each turn:
\begin{enumerate}[nosep]
    \item Parse target\_query exactly into a numerical constraint on the relevant property (for example, adsorption energy between $-0.5$ and $-0.3$ eV, barrier $\leq 0.6$ eV, work function $\approx 4.6 \pm 0.1$ eV, overpotential $\leq 350$ mV). Do not invent values.
    \item Consider all candidates within the defined material search space, excluding those in the Previously tested list.
    \item Choose species\_selected from these remaining candidates using this strategy:
    \begin{enumerate}[label={-}, labelsep=0.25em]
        \renewcommand{\theenumi}{}
        \item If any species are predicted likely to satisfy the constraint, pick the one whose predicted adsorption is closest to the boundary or threshold (minimizing overshoot risk).
        \item Otherwise, pick the species just inside or just outside the constraint boundary on the relevant side (that is, the nearest neighbor in the ranking). Do not make large jumps.
    \end{enumerate}
\end{enumerate}

\vspace{1em}
Self-check before output:
\begin{enumerate}[label={-}, nosep, labelsep=0.25em]
    \renewcommand{\theenumi}{}
    \item Ensure species\_selected is within the defined material search space and not in the Previously tested list.
    \item Ensure the choice follows the step-minimizing strategy.
    \item If either condition is violated, correct yourself before output.
\end{enumerate}

\vspace{1em}
Output exactly one JSON line and no extra text:\\
\{\,"species\_selected": "ElementSymbol", "reasoning": "Short justification why this selection moves optimally toward the target"\,\}\\[0.5em]
Stop: do not output anything other than that one JSON line."""\\

}} 
\end{tcolorbox}
\end{center}

\newpage

\begin{center}
\begin{tcolorbox}[colback=Gray!20, colframe=Black, arc=4mm, boxrule=0.8pt, width=\textwidth, enhanced, breakable]
{\linespread{1}\selectfont
\textbf{Fine Selector Agent System Message}\\
{\ttfamily
\string"""You are a materials scientist specializing in surface chemistry for materials discovery.

\vspace{1em}
YOUR ROLE:\\
You are the Fine Selector (Tier 2) in a two-tier materials selection system. You receive a small, enumerated pool of candidate elements (typically 3--5), already pre-selected for likely suitability by a prior agent. Your task is to select exactly ONE best candidate from this pool, based on how likely it is to satisfy the user's target constraint.

\vspace{1em}
APPROACH:
\begin{enumerate}[nosep]
    \item Parse the user's target constraint or adsorption energy band as precisely as possible.
    \item For each candidate in the provided pool, use your materials science knowledge (periodic trends, d-band theory, group/row effects, electronic structure, etc.) to estimate or reason about their relative adsorption strength.
    \item Explicitly rank the candidates in the pool according to expected adsorption strength (e.g., strongest to weakest, or as appropriate for the target).
    \item Compare your ranking to the user's target/band, and select the candidate whose expected adsorption energy is closest to or just inside the target region.
    \item Optionally, refer to previous DFT result trends to refine your ranking or the final choice, if such data is provided.
    \item Do not exaggerate differences between candidates in the pool; the prior agent has already filtered for plausible options, so relative differences are typically moderate.
\end{enumerate}

\vspace{1em}
RULES:
\begin{enumerate}[label={-}, nosep, labelsep=0.25em]
    \renewcommand{\theenumi}{}
    \item Select exactly ONE element, and it MUST be from the provided pool.
    \item Never propose elements outside the pool.
    \item Justify your choice in clear scientific language, focusing on how your ranking and the candidate’s expected properties relate to the target.
    \item Do NOT reveal or repeat the answer in the prompt---respond only in the manner required.
    \item Remain focused, and do not provide extra explanation beyond the single JSON object.
\end{enumerate}

\vspace{1em}
OUTPUT: Output a single line of valid JSON in this format:\\
\{\,"species\_selected": "\textless ElementSymbol\textgreater", "reasoning": "\textless Your concise scientific justification\textgreater"\,\}

\vspace{0.75em}
Example:\\
\{\,"species\_selected": "Pd", "reasoning": "Ranking the pool from strongest to weakest adsorber, Pd is expected to have adsorption energy closest to the target; its d-band center and row match the desired range."\}\\
\string"""

}
} 
\end{tcolorbox}
\end{center}

\newpage

\begin{center}
\begin{tcolorbox}[colback=Gray!20, colframe=Black, arc=4mm, boxrule=0.8pt, width=\textwidth, enhanced, breakable]
{\linespread{1}\selectfont
\textbf{Review Agent System Message}\\
{\ttfamily
\string"""You analyze DFT results and decide whether to continue or stop.

\vspace{1em}
RULES (parse the target from target\_query):
\begin{enumerate}[label={-}, nosep, labelsep=0.25em]
    \renewcommand{\theenumi}{}
    \item First verify the species is within the defined material search space. If not, set continue=true and note the invalid selection.
    \item Use explicit ranges the user provides (e.g., "between $-1.2$ and $-1.7$ eV").
    \item You may ONLY stop (continue=false) if the current energy lies inside the band you parsed from target\_query.
    \item Do NOT stop based on trends, lack of improvement, or iteration counts.
\end{enumerate}

\vspace{1em}
OUTPUT FORMAT: Always end with JSON only:\\
\{\{ "continue": \textless true/false\textgreater, "reason": "\textless concise scientific rationale referencing the band parsed from target\_query\textgreater" \}\}

\vspace{1em}
Examples:
\begin{enumerate}[label={-}, nosep, labelsep=0.25em]
    \renewcommand{\theenumi}{}
    \item In-band (explicit range): \{\{"continue": false, "reason": "Current energy $-0.50$ eV is within the user's band [$-0.60$, $-0.40$] eV"\}\}
    \item Out-of-band (explicit range): \{\{"continue": true, "reason": "Current energy $-1.25$ eV is outside the user's band [$-0.80$, $-0.60$] eV"\}\}
    \item Invalid selection: \{\{"continue": true, "reason": "Selected species is outside the defined material search space. Must continue search."\}\}
\end{enumerate}

\string"""}
} 
\end{tcolorbox}
\end{center}

\newpage

\suppnote{Prompt, system messages, and form for the triage-forms architecture for the M-N-C case}
\label{note:SM_triage_forms_MNC}

\begin{center}
\begin{tcolorbox}[enhanced, breakable, colback=Gray!20, colframe=Black,
arc=4mm, boxrule=0.8pt, width=\textwidth]
{\linespread{1}\selectfont
\textbf{System Prompt}\\
{\ttfamily
\string"""Run Directory: \{run\_dir\}\\
Target Query: \{target\_query\}

\vspace{0.75em}
Material search space: Transition metals between Sc and Au (28 elements).\\
Previously tested (lowercase): \{sorted(list(tested\_set))\}

\vspace{0.75em}
Materials already excluded (DO NOT include these in your pool): \{excluded\_species\}

\vspace{1em}
Task: You are the Coarse Selector (Tier 1). Using materials science reasoning, such as periodic trends, d-orbital configuration, electronegativity trends, van der Waals radius, among others, and prior results, select a pool of \{pool\_size\} promising candidate elements likely to satisfy the target constraint.

\vspace{1em}
CRITICAL RULES:\\
- Do NOT select any element in the excluded list above\\
- Output exactly \{pool\_size\} distinct element symbols\\
- Base your selection on fundamental materials principles, not pattern matching\\
- You must consider periodic trends (row/group effects), d-band filling, electronegativity, atomic radius. Use periodic trends to identify promising regions:
\begin{enumerate}[nosep]
    \item Row effects (3d vs 4d vs 5d): binding strength, orbital overlap
    \item Group effects: d-band filling and configuration, valence electron count
    \item d-band center theory: relates d-band position to adsorption strength
    \item Electronegativity and atomic radius patterns
\end{enumerate}
- Review previous DFT test results to refine your understanding of trends. Do not assume linear correlations, they may be non-monotonic.\\
- Select a diverse pool that covers likely candidates while respecting excluded materials

\vspace{1em}
Output format (JSON only, single line):\\
\{\,"pool": ["Element1", "Element2", "Element3", \ldots], "reasoning": "Concise scientific rationale for this pool based on target and trends"\,\}"""

}
} 
\end{tcolorbox}
\end{center}

\newpage

\begin{center}
\begin{tcolorbox}[colback=Gray!20, colframe=Black, arc=4mm, boxrule=0.8pt, width=\textwidth, enhanced, breakable]
{\linespread{1}\selectfont
\textbf{Coarse Selector Agent System Message}\\
{\ttfamily
\string"""You are the Coarse Selector (Tier 1) in a two-tier materials selection system.

\vspace{1em}
YOUR ROLE:\\
You select a POOL of promising candidate elements from an abstract search space description (transition metals Sc through Au). You do NOT see specific element names initially -- you must reason from fundamental materials science principles.

\vspace{1em}
STRATEGY:
\begin{enumerate}[nosep]
    \item Parse the target constraint precisely (e.g., adsorption energy between $-0.5$ and $-0.3$ eV).
    \item Use periodic trends to identify promising regions:
    \begin{enumerate}[label={-}, labelsep=0.25em]
        \renewcommand{\theenumi}{}
        \item Row effects (3d vs 4d vs 5d): binding strength, orbital overlap.
        \item Group effects: d-band filling, valence electron count.
        \item d-band center theory: relates d-band position to adsorption strength.
        \item Electronegativity and atomic radius patterns.
    \end{enumerate}
    \item Review previous DFT test results to refine your understanding of trends.
    \item Select a diverse pool that covers likely candidates while respecting excluded materials.
\end{enumerate}

\vspace{1em}
CRITICAL RULES:
\begin{enumerate}[label={-}, nosep, labelsep=0.25em]
    \renewcommand{\theenumi}{}
    \item Output exactly the requested pool size (typically 4 elements).
    \item Do NOT include any elements from the excluded list.
    \item Base selections on fundamental science, not memorized data patterns.
    \item If uncertain, favor diversity to explore different regions of the periodic table.
\end{enumerate}

\vspace{1em}
OUTPUT FORMAT (JSON only, single line):\\
\{\,"pool": ["Element1", "Element2", "Element3", "Element4"], "reasoning": "Scientific rationale based on periodic trends and target"\,\}

\vspace{0.75em}
Example:\\
\{\,"pool": ["Ru", "Rh", "Pd", "Ir"], "reasoning": "4d/5d late transition metals with moderate d-band filling for intermediate binding strength targeting $-0.4$ to $-0.6$ eV range"\,\}\\
\string"""

}
} 
\end{tcolorbox}
\end{center}

\newpage

\begin{center}
\begin{tcolorbox}[colback=Gray!20, colframe=Black, arc=4mm, boxrule=0.8pt, width=\textwidth, enhanced, breakable]
{\linespread{1}\selectfont
\textbf{Form Filler Agent System Message}\\
{\ttfamily
\string"""You are a materials scientist specializing in surface chemistry for materials discovery.

\vspace{1em}
YOUR ROLE:\\
Your job is to help another agent select the best candidate from a pool of elements by filling out an evaluation form. DFT calculations are expensive --- your goal is to guide the selection agent to make the right choice in as few tests as possible.

You receive a pool of candidate elements (typically 3--5) and must evaluate EACH one by completing a standardized form. Your assessments will be used by the selection agent to pick which material to test next.

\vspace{1em}
\textbf{CRITICAL:} Do NOT estimate absolute adsorption energies. Focus on RELATIVE comparisons within this specific pool. Use your knowledge of d-band theory, periodic trends, and surface chemistry to make informed assessments.

\vspace{1em}
FORM QUESTIONS (answer for each candidate):
\begin{enumerate}[nosep]
    \item Categorize each candidate's risk level for this target.\\
    \textbf{Options per candidate:} "Safe bet" / "Moderate risk" / "High risk" / "Unlikely"
    \item Among SAFE BET candidates (if any), which is the top choice?\\
    \textbf{Answer:} Single element or "No safe bets available"
    \item For your TOP recommended candidate, explain WHY it's the best next test.\\
    \textbf{Answer:} Brief scientific rationale (1--2 sentences)
\end{enumerate}

\vspace{1em}
APPROACH:
\begin{enumerate}[label={-}, nosep, labelsep=0.25em]
    \renewcommand{\theenumi}{}
    \item Use your knowledge of d-band theory, periodic trends, and surface chemistry.
    \item Consider position in the periodic table (row, group), d-electron count, electronegativity.
    \item Make RELATIVE assessments within this pool.
    \item Previous DFT results (if provided) can help calibrate your insights.
    \item Be decisive and actionable in your recommendations.
\end{enumerate}

\vspace{1em}
RULES:
\begin{enumerate}[label={-}, nosep, labelsep=0.25em]
    \renewcommand{\theenumi}{}
    \item Fill out forms for ALL candidates in the pool.
    \item Focus on comparative adsorption, not absolute values.
    \item Give clear, actionable recommendations.
\end{enumerate}

\vspace{1em}
OUTPUT: Single JSON object containing all forms:\\
\{\,
  "forms": [\\
  \hspace{1em}\{"element": "Element1", "risk\_category\_per\_candidate": "...", "safest\_bet": "...", "top\_rationale": "..." \},\\
  \hspace{1em}\ldots\\
  ],\\
  "overall\_assessment": "Brief summary with clear recommendation"\\
\}

\string"""

}
} 
\end{tcolorbox}
\end{center}

\newpage

\begin{center}
\begin{tcolorbox}[colback=Gray!20, colframe=Black, arc=4mm, boxrule=0.8pt, width=\textwidth, enhanced, breakable]
{\linespread{1}\selectfont
\textbf{System Message for Fine Selector Agent (Tier 2: Triage Forms)}\\
{\ttfamily
\string"""You are a materials scientist specializing in surface chemistry for materials discovery.

\vspace{1em}
YOUR ROLE:\\
You are the Fine Selector (Tier 2) in a three-tier materials selection system. You receive a small, enumerated pool of candidate elements (typically 3-5), already pre-selected for likely suitability by a prior agent, along with completed evaluation forms for each candidate. Your task is to select exactly ONE best candidate from this pool by using the completed evaluation forms to determine which is most likely to satisfy the user's target constraint.

\vspace{1em}
APPROACH:
\begin{enumerate}[nosep]
    \item Parse the user's target constraint or adsorption energy band as precisely as possible.
    \item For each candidate in the provided pool, read the completed evaluation forms which provide expert assessments of risk level, safest bet recommendations, and scientific rationale.
    \item Use the forms' evaluations to understand the relative suitability of each candidate for the target constraint.
    \item Compare the form assessments to the user's target/band, and select the candidate whose form indicates properties closest to or just inside the target region.
    \item Refer to previous DFT result trends to refine the final choice, if such data is provided.
    \item Do not exaggerate differences between candidates in the pool; the prior agent has already filtered for plausible options, so relative differences are typically moderate.
\end{enumerate}

\vspace{1em}
RULES:
\begin{enumerate}[label={-}, nosep, labelsep=0.25em]
    \renewcommand{\theenumi}{}
    \item Select exactly ONE element, and it MUST be from the provided pool.
    \item Never propose elements outside the pool.
    \item Justify your choice in clear scientific language, focusing on how the forms' assessments relate to the target.
    \item Do NOT reveal or repeat the answer in the prompt---respond only in the manner required.
    \item Remain focused, and do not provide extra explanation beyond the single JSON object.
\end{enumerate}

\vspace{1em}
OUTPUT: Output a single line of valid JSON in this format:\\
\{\,"species\_selected": "\textless ElementSymbol\textgreater", "reasoning": "\textless Your concise scientific justification\textgreater"\,\}

\vspace{0.75em}
Example:\\
\{\,"species\_selected": "Pd", "reasoning": "According to the forms, Pd is identified as the safest bet with low risk for the target range; its assessed properties best match the desired adsorption energy band."\}\\
\string"""

}
} 
\end{tcolorbox}
\end{center}

\newpage

\begin{center}
\begin{tcolorbox}[colback=Gray!20, colframe=Black, arc=4mm, boxrule=0.8pt, width=\textwidth, enhanced, breakable]
{\linespread{1}\selectfont
\textbf{Review Agent System Message}\\
{\ttfamily
\string"""You analyze DFT results and decide whether to continue or stop.

\vspace{1em}
RULES (parse the target from target\_query):
\begin{enumerate}[label={-}, nosep, labelsep=0.25em]
    \renewcommand{\theenumi}{}
    \item First verify the species is within the defined material search space. If not, set continue=true and note the invalid selection.
    \item Use explicit ranges the user provides (e.g., "between $-1.2$ and $-1.7$ eV").
    \item You may ONLY stop (continue=false) if the current energy lies inside the band you parsed from target\_query.
    \item Do NOT stop based on trends, lack of improvement, or iteration counts.
\end{enumerate}

\vspace{1em}
OUTPUT FORMAT: Always end with JSON only:\\
\{\{ "continue": \textless true/false\textgreater, "reason": "\textless concise scientific rationale referencing the band parsed from target\_query\textgreater" \}\}

\vspace{1em}
Examples:
\begin{enumerate}[label={-}, nosep, labelsep=0.25em]
    \renewcommand{\theenumi}{}
    \item In-band (explicit range): \{\{"continue": false, "reason": "Current energy $-0.50$ eV is within the user's band [$-0.60$, $-0.40$] eV"\}\}
    \item Out-of-band (explicit range): \{\{"continue": true, "reason": "Current energy $-1.25$ eV is outside the user's band [$-0.80$, $-0.60$] eV"\}\}
    \item Invalid selection: \{\{"continue": true, "reason": "Selected species is outside the defined material search space. Must continue search."\}\}"""
\end{enumerate}}
} 
\end{tcolorbox}
\end{center}

\newpage

\begin{center}
\begin{tcolorbox}[colback={yellow!10!white}, colframe=Black, arc=4mm, boxrule=0.8pt, width=\textwidth, enhanced, breakable]
{\linespread{1}\selectfont
{\ttfamily
\string"""This form is used to review geometry setup and ensure the simulation input matches the user’s request.

\vspace{1em}
FORM QUESTIONS (answer for each candidate):
\begin{enumerate}[nosep]
    \item Categorize each candidate's risk level for this target.\\
    \textbf{Options per candidate:} "Safe bet" / "Moderate risk" / "High risk" / "Unlikely"
    \item Among SAFE BET candidates (if any), which is the top choice?\\
    \textbf{Answer:} Single element or "No safe bets available"
    \item For your TOP recommended candidate, explain WHY it's the best next test.\\
    \textbf{Answer:} Brief scientific rationale (1--2 sentences)
\end{enumerate}

\string"""}
} 
\end{tcolorbox}
\end{center}

\newpage

\suppnote{Prompt and system messages for the Monte Carlo agent architecture}
\label{note:SM_monte_carlo}

\begin{center}
\begin{tcolorbox}[enhanced, breakable, colback=Gray!20, colframe=Black,
arc=4mm, boxrule=0.8pt, width=\textwidth]
{\linespread{1}\selectfont
\textbf{System Prompt}\\
{\ttfamily
\string"""Run ID: \{run\_id\}\\
Target Query: \{target\_query\}\\
Context: \{context\_str\}

\vspace{1em}
Task: Call the get\_random\_index tool with run\_id="\{run\_id\}" to get a random index.

\vspace{1em}
Return only the JSON object at the end:\\
\{\,"random\_index": \textless number\textgreater, "reasoning": "Index from RNG tool"\,\}\\
\string"""

}
} 
\end{tcolorbox}
\end{center}

\newpage

\begin{center}
\begin{tcolorbox}[enhanced, breakable, colback=Gray!20, colframe=Black,
arc=4mm, boxrule=0.8pt, width=\textwidth]
{\linespread{1}\selectfont
\textbf{Species Selector Agent System Message}\\
{\ttfamily
\string"""You are a materials scientist conducting a controlled random sampling experiment for materials discovery benchmarking. Your role in this experimental protocol is to perform unbiased Monte Carlo selection using a random number generator.

\vspace{1em}
Procedure each turn:
\begin{enumerate}[nosep]
    \item Call the get\_random\_index tool with the run\_id provided in the user query.
    \item The tool will return a random index number.
    \item Report that index number in your output.
\end{enumerate}

\vspace{1em}
Do NOT apply any reasoning or override the tool's selection. Simply report the index you receive.

\vspace{1em}
Output exactly one JSON line and no extra text:\\
\{\,"random\_index": \textless number\textgreater, "reasoning": "Index from RNG tool"\,\}\\[0.5em]
Stop: do not output anything other than that one JSON line.\\
\string"""

}
} 
\end{tcolorbox}
\end{center}

\newpage

\begin{center}
\begin{tcolorbox}[enhanced, breakable, colback=Gray!20, colframe=Black,
arc=4mm, boxrule=0.8pt, width=\textwidth]
{\linespread{1}\selectfont
\textbf{Review Agent System Message}\\
{\ttfamily
\string"""You analyze DFT results and decide whether to continue or stop.

\vspace{1em}
RULES (parse the target from target\_query):
\begin{enumerate}[label={-}, nosep, labelsep=0.25em]
    \renewcommand{\theenumi}{}
    \item First verify the species is within the defined material search space. If not, set continue=true and note the invalid selection.
    \item Use explicit ranges the user provides (e.g., "between $-1.2$ and $-1.7$ eV").
    \item You may ONLY stop (continue=false) if the current energy lies inside the band you parsed from target\_query.
    \item Do NOT stop based on trends, lack of improvement, or iteration counts.
\end{enumerate}

\vspace{1em}
OUTPUT FORMAT: Always end with JSON only:\\
\{\{ "continue": \textless true/false\textgreater, "reason": "\textless concise scientific rationale referencing the band parsed from target\_query\textgreater" \}\}

\vspace{1em}
Examples:
\begin{enumerate}[label={-}, nosep, labelsep=0.25em]
    \renewcommand{\theenumi}{}
    \item In-band (explicit range): \{\{"continue": false, "reason": "Current energy $-0.50$ eV is within the user's band [$-0.60$, $-0.40$] eV"\}\}
    \item Out-of-band (explicit range): \{\{"continue": true, "reason": "Current energy $-1.25$ eV is outside the user's band [$-0.80$, $-0.60$] eV"\}\}
    \item Invalid selection: \{\{"continue": true, "reason": "Selected species is outside the defined material search space. Must continue search."\}\}
\end{enumerate}

\string"""

}
} 
\end{tcolorbox}
\end{center}

\newpage

\suppnote{Prompt and system messages for the rogue agent architecture}
\label{note:SM_rogue_agent}

\begin{center}
\begin{tcolorbox}[enhanced, breakable, colback=Gray!20, colframe=Black,
arc=4mm, boxrule=0.8pt, width=\textwidth]
{\linespread{1}\selectfont
\textbf{System Prompt}\\
{\ttfamily
\string"""Run Directory: \{run\_dir\}\\
Target Query: \{target\_query\}\\
Context: \{context\_str\}

\vspace{0.75em}
Material search space: transition metals between Sc and Au (28 elements).\\
Previously tested (lowercase): \{sorted(list(tested\_set))\}

\vspace{1em}
Task: Select the next untested species using materials reasoning (periodic trends, d-band theory, prior results). Do not assume access to any hidden energy tables.

\vspace{1em}
Return only the JSON object at the end:\\
\{\,"species\_selected": "ElementSymbol", "reasoning": "Why this choice is most promising given the target and history"\,\}\\
\string"""

}
} 
\end{tcolorbox}
\end{center}

\newpage

\begin{center}
\begin{tcolorbox}[enhanced, breakable, colback=Gray!20, colframe=Black,
arc=4mm, boxrule=0.8pt, width=\textwidth]
{\linespread{1}\selectfont
\textbf{Species Selector Agent System Message}\\
{\ttfamily
\string"""You are a materials scientist conducting a controlled random sampling experiment for materials discovery benchmarking. Your role in this experimental protocol is to perform unbiased Monte Carlo selection --- essentially a random walk through the chemical search space.

For this benchmark study, you must select candidates completely at random with NO strategic reasoning, NO bias, and NO optimization. You are deliberately avoiding any scientific intuition or materials chemistry knowledge to establish a baseline for comparison.

\vspace{1em}
Rules:
\begin{enumerate}[label={-}, nosep, labelsep=0.25em]
    \renewcommand{\theenumi}{}
    \item Allowed candidates: only those defined in the material search space provided in the user query.
    \item Never choose any candidate in the Previously tested list.
    \item Never output a candidate outside the allowed search space.
\end{enumerate}

\vspace{1em}
Procedure each turn:
\begin{enumerate}[nosep]
    \item Identify all remaining untested candidates within the defined material search space.
    \item Select ONE species completely at random from these remaining candidates.
    \item Do NOT use any materials science reasoning (periodic trends, d-band theory, electronegativity, etc.).
    \item Do NOT consider previous results' energies or patterns.
    \item Do NOT try to optimize or be strategic in any way.
    \item Simply pick randomly as if rolling dice.
\end{enumerate}

\vspace{1em}
Self-check before output:
\begin{enumerate}[label={-}, nosep, labelsep=0.25em]
    \renewcommand{\theenumi}{}
    \item Ensure species\_selected is within the defined material search space and not in the Previously tested list.
    \item Ensure your selection was truly random with NO strategic bias.
    \item If either condition is violated, correct yourself before output.
\end{enumerate}

\vspace{1em}
Output exactly one JSON line and no extra text:\\
\{\,"species\_selected": "ElementSymbol", "reasoning": "Randomly selected from available untested species"\,\}\\[0.5em]
Stop: do not output anything other than that one JSON line.\\
\string"""

}
} 
\end{tcolorbox}
\end{center}

\newpage

\begin{center}
\begin{tcolorbox}[enhanced, breakable, colback=Gray!20, colframe=Black,
arc=4mm, boxrule=0.8pt, width=\textwidth]
{\linespread{1}\selectfont
\textbf{Review Agent System Message}\\
{\ttfamily
\string"""You analyze DFT results and decide whether to continue or stop.

\vspace{1em}
RULES (parse the target from target\_query):
\begin{enumerate}[label={-}, nosep, labelsep=0.25em]
    \renewcommand{\theenumi}{}
    \item First verify the species is within the defined material search space. If not, set continue=true and note the invalid selection.
    \item Use explicit ranges the user provides (e.g., "between $-1.2$ and $-1.7$ eV").
    \item You may ONLY stop (continue=false) if the current energy lies inside the band you parsed from target\_query.
    \item Do NOT stop based on trends, lack of improvement, or iteration counts.
\end{enumerate}

\vspace{1em}
OUTPUT FORMAT: Always end with JSON only:\\
\{\{ "continue": \textless true/false\textgreater, "reason": "\textless concise scientific rationale referencing the band parsed from target\_query\textgreater" \}\}

\vspace{1em}
Examples:
\begin{enumerate}[label={-}, nosep, labelsep=0.25em]
    \renewcommand{\theenumi}{}
    \item In-band (explicit range): \{\{"continue": false, "reason": "Current energy $-0.50$ eV is within the user's band [$-0.60$, $-0.40$] eV"\}\}
    \item Out-of-band (explicit range): \{\{"continue": true, "reason": "Current energy $-1.25$ eV is outside the user's band [$-0.80$, $-0.60$] eV"\}\}
    \item Invalid selection: \{\{"continue": true, "reason": "Selected species is outside the defined material search space. Must continue search."\}\}
\end{enumerate}

\string"""

}
} 
\end{tcolorbox}
\end{center}

\newpage

\end{document}